\documentclass[12pt]{article}
\usepackage{ifpdf}
\usepackage{amsmath,amsthm,amsfonts,amssymb,amscd}
\usepackage{wasysym}
\usepackage{graphicx}
\usepackage{wrapfig}
\usepackage{multirow}
\usepackage{longtable}
\usepackage{array}

\newcommand{\nd} {\noindent}

\begin{document}
\begin{titlepage}

\begin{center}

  {\LARGE \bf A power law decay evolution scenario for polluted single white dwarfs}

\end{center}
 
\vskip  1.0truecm
\centerline {{\large Di-Chang Chen, Ji-Lin Zhou$^{*}$, Ji-Wei Xie$^{*}$, Ming Yang, Hui Zhang, }}
\centerline {{\large Hui-Gen Liu, En-Si Liang, Zhou-Yi Yu,  Jia-Yi Yang}}

\vskip 0.5truecm

\centerline { {\sl School of Astronomy and Space Science, Key Laboratory of Ministry of Education,  }}
\centerline {{\sl Nanjing University, Nanjing 210046, China.}}

\vskip .5truecm

\begin{abstract} 
Planetary systems can survive the stellar evolution, as evidenced
by the atmospheric metal pollution and circumstellar dusty disks of single white dwarfs\cite{Koester2014,Farihi2016a}. Recent observations show that $1\%-4\%$ of single white dwarfs are accompanied by dusty disks\cite{Mullally2007,Farihi2009,Debes2011a,Barber2014}, while the occurrence rate of metal pollution is about $25\%-50\%$\cite{Koester2014,Zuckerman2003,Zuckerman2010}. The dusty disks and metal pollution have been associated with accretion of remanent planetary systems around white dwarfs\cite{Koester2014,Bonsor2017}, yet the relation between these two phenomena is still unclear.
Here we suggest an evolutionary scenario to link the two observational phenomena.
By analyzing a sample of metal polluted white dwarfs, we find that the mass accretion rate onto the white dwarf generally follows a broken power law decay, which matches well with the theoretical prediction, if assuming dust accretion is primarily driven by Poynting-Robertson drag \cite{Rafikov2011a} and the dust source is primarily delivered via dynamically falling asteroids perturbed by a Jovian planet\cite{Frewen2014}.    
The presence of disks is mainly at the early stage ($t_{\rm cool}\sim 0.1-0.7$ Gyr) of the whole process of metal pollution, which is detectable until $\thicksim$ 8 Gyr, naturally explaining the fraction ($\sim 2\%$ -- $16\%$) of metal-polluted white dwarfs having dusty disks. The success of this scenario also implies that the configuration of an asteroid belt with an outer gas giant might be common around stars of several solar masses.
\end{abstract} 

\vskip .5truecm

\noindent



\vfill
\hrule
\smallskip

\nd  e-mail: zhoujl@nju.edu.cn;  jwxie@nju.edu.cn
\end{titlepage}

\pagebreak

White dwarf stars (WDs) are the final evolution stage of main sequence stars with masses less than about $8M_{\bigodot}$ (solar mass)\cite{Weidemann2000}. During the post main sequence evoultion, the envelope of such a main sequence star will expand to several AU, therefore planets in the inner region might be swallowed, while the orbits of outer planets would expand due to stellar mass loss\cite{Mustill2012}. Thus the study of environment around WDs can reveal the structure of exoplanet systems around the progenitor stars.

In the last two decades, among thousands of single WDs with effective temperature below 25,000K, hundreds of metal-polluted ones have been identified, indicating ongoing or recent accretion of remanent planetary systems beyond several AU\cite{Koester2014}. About 40 of them show infrared excesses in their spectra, indicating the presence of circumstellar dusty disks\cite{Farihi2016a}. 

It is not clear yet what physical mechanisms cause the difference between the occurrence rates of disk-possessing WDs and metal-polluted ones. The reason why the occurrence rate of infrared detectable disk decreases with age on a timescale of roughly 0.5 Gyr\cite{Farihi2009} is not well understood, either. By analyzing a sample of metal polluted WDs and combing with theoretical calculations as well as dynamical simulations,  we find that these questions can be answered self-consistently in a unified evolution scenario.

\section{Result}

\subsection*{Power law decay derived from observation data}
We collect an observational sample of metal-polluted WDs from literatures based on the Montreal White Dwarf Database (MWDD)\cite{Dufour2017}.
Our sample consists of 47 hydrogen-atmosphere (DA) and 799 helium-atmosphere (non-DA) WDs.
We perform a uniform analysis (section 1 of Methods) through the sample to calculate the mass accretion rate ($\dot{M}_{\rm Z}$) and the cooling age ($t_{\rm cool}$). 

Figure 1a shows the accretion rate as a function of the cooling age for the whole sample.
As can be clearly seen,  $\dot{M}_{\rm Z}$  decays significantly with increasing $t_{\rm cool}$.
 We fit the data with a constant model, a single power law model and a broken power law decay respectively  (section 2 of Methods).
The single power law decay model obtains a much smaller Akaike Information Criterion (AIC) score ($\Delta \rm AIC= 527$) than that of the constant model (Supplementary Figure 2), indicating the former model is much better.
Nevertheless, we note that the observations apparently deviates from the single power law model and breaks up into two branches at early phase  (i.e., $t_{\rm cool}<0.3$ Gyr).  
The shallow branch is dominated by DA WDs with $\dot{M}_{\rm Z}$ always less than $2\times10^9\rm gs^{-1}$. 
The steep branch is dominated by non-DA WDs, whose $\dot{M}_{\rm Z}$ can be as high as $10^{11}\rm gs^{-1}$.
Such a discrepancy is not unexpected because $\dot{M}_{\rm Z}$ derived from DA and no-DA WDs are actually not exactly the same.
DA WDs usually have short metal sinking timescale, and the derived $\dot{M}_{\rm Z}$ can be inferred as the currently on-going mass accretion rate.
Non-DA WDs usually have much longer sinking timescale (section 4.4 of Methods), and the derived $\dot{M}_{\rm Z}$ only reflects the average of historical accretion rate, which could be severely dominated by some relatively rare outliers, e.g., runaway accretion \cite{Rafikov2011b} with abnormally high accretion rate.
Therefore, we expect that the derived $\dot{M}_{\rm Z}$ from DA (rather than Non-DA) WDs to be a better tracer to reveal the baseline of the $\dot{M}_{\rm Z}$ evolution.

We then focus only on the DA WDs.
Figure 1b shows $\dot{M}_{\rm Z}$ as a function of $t_{\rm cool}$ for only DA WDs in the sample.  
Again, a constant accretion rate model can be confidently ruled out as compared to a power law decay model with an AIC score difference $\Delta \rm AIC= 26$. 
Furthermore, a broken power law model is marginally preferred than a single power law model with $\Delta \rm AIC = 2$ (Supplementary Figure 4). 
The best fit model is denoted as the blue broken line in Figure 1b and mathematically given by  
\begin{equation}
\log\left(\frac{\dot{M}_{\rm Z}}{\rm gs^{-1}}\right) =
\begin{cases}
{\rm -0.72}\times \log\left(\frac{t_{\rm cool}}{\rm Gyr}\right)+ {\rm 7.81},& \text{$t_{\rm cool}<0.68 \rm \ Gyr$}\\
{\rm -2.07}\times \log\left(\frac{t_{\rm cool}}{\rm Gyr}\right)+ {\rm 7.58},& \text{$t_{\rm cool}\ge 0.68 \rm \ Gyr$}
\end{cases}
\label{dotMz}
\end{equation}
  
The above derived power law decay of $\dot{M}_{\rm Z}$ from observation is unlikely to be caused by observational bias (see section 4.1 of Methods).
Instead, it matches well with the theoretical predictions as we show below. 

\begin{figure*}
\centering
\includegraphics[width=14cm]{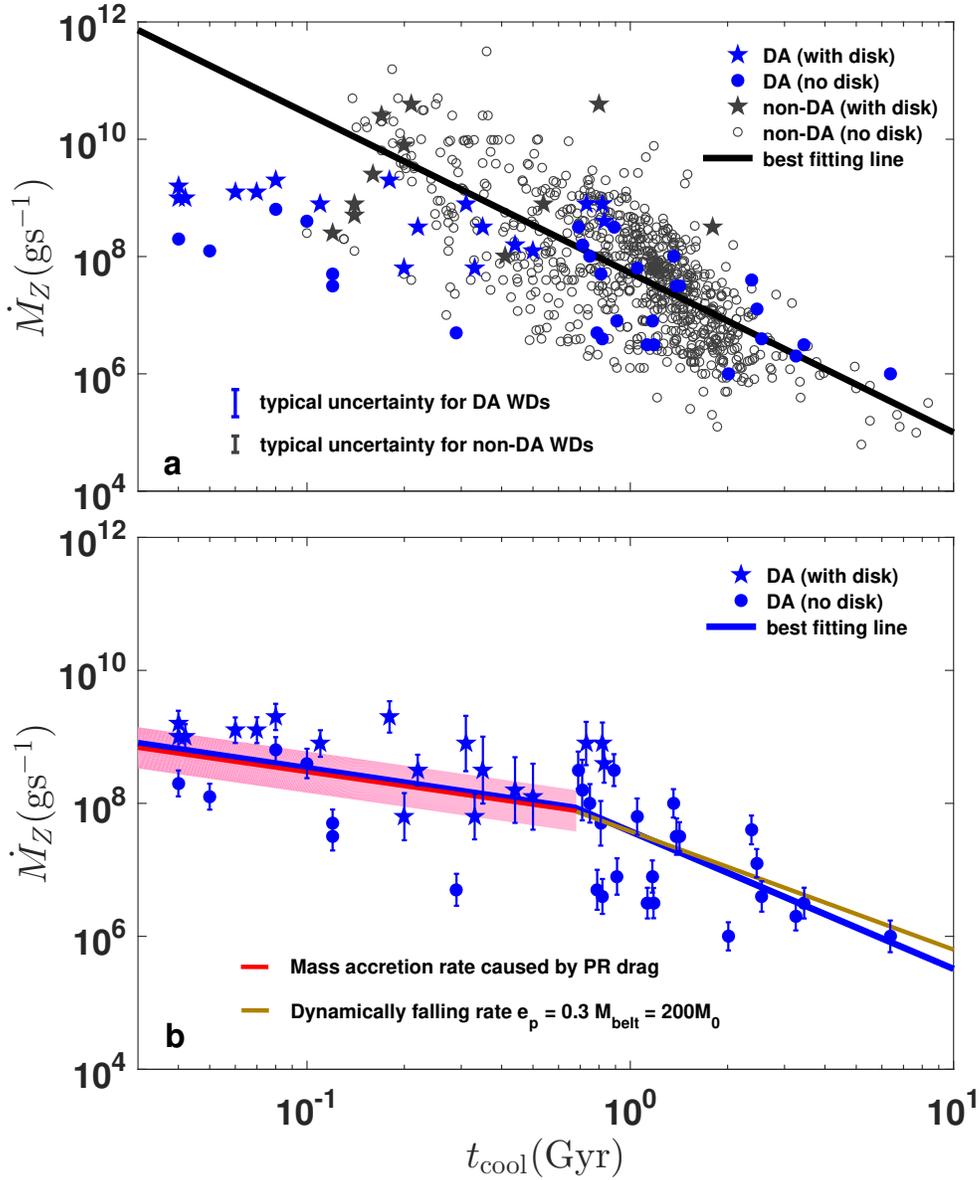}
\caption{ {\bfseries The power law relationship derived from the observational data.}  {\bfseries a.} The accretion rate $\dot{M}_{\rm Z}$ is plotted as a function of cooling age $t_{\rm cool}$ for the whole sample. The solid black line denotes the best fitting line of $\dot{M}_{\rm Z}$ and $t_{\rm cool}$. The typical (median) uncertainties in $\dot{M}_{\rm Z}$ for DA subsample (blue) and non-DA subsample (grey) are shown in the bottom left corner of the panel. {\bfseries b.} $\dot{M}_{\rm Z}$ is plotted as a function of $t_{\rm cool}$ for the DA subsample. The best fit for the DA subsample (blue solid line) is a broken power law, which matches well with the accretion rate caused by PR (red region for 1 $\sigma$ range and solid red line for a typical case, Eqn.\ref{MPRt}) followed by the dynamically falling rate (solid brown line) of a typical case with a planet eccentricity of 0.3 and asteroid belt mass of 200 times the main asteroid belt. }
\label{fig1}
\end{figure*}

\subsection*{Power law decay derived from theoretical models}
Based on the standard tidally disrupted asteroid model\cite{Farihi2016a,Jura2003}, assuming a steady accretion phase for an opaque disk with a narrow optical transition zone, we derive the relationship between the mass accretion rate caused by the Poynting-Robertson (PR) drag \cite{Rafikov2011a} and the cooling age (see section 3.1 of Methods), which is a power law decay as 
\begin{equation}
\log\left(\frac {\dot{M}_{\rm PR}}{\rm gs^{-1}}\right)=-0.70\times \log\left(\frac{t_{\rm cool}}{\rm Gyr}\right)+7.77(\pm 0.31)
\label{MPRt}
\end{equation}
Interestingly, this analytical power law decay due to PR effect is almost identical to the early stage of the observational best fit (Eqn. \ref{dotMz}). 
Such a good match reinforces two expectations.
First, PR effect is a primary mechanism to drive mass accretion in the early phase.
Second, $\dot{M}_{\rm Z}$ derived from DA (other than non-DA) WDs are reliable to trace the baseline of mass accretion evolution.

In order to maintain mass accretion on to WDs by PR effect, material should be continually delivered to the star. 
Various source delivering models have been proposed \cite{Frewen2014, Bonsor2011, Bonsor2015, Hamers2016, Payne2016}.
The essence of many of these models is that small bodies, e.g., asteroid analog, moon analog, crash into the WD host star due to perturbations of other big bodies, e.g., Jupiter analog.  
Here we consider a simplified solar system remnant as a fiducial model (see section 3.2 of Methods), which consists of a central WD, an asteroid belt and a Jupiter-mass planet. 
Due to the planetary perturbations, the orbits of asteroids become chaotic, reaching very large orbital eccentricity with periastron within the Roche limit of the central WD.
We record the flux of the asteroids dynamically falling into the Roche limit ($\dot{M}_{\rm DF}$), and find it fits well with a power law function of time.
This is not unexpected because for a typical Hamiltonian system, particles escaping from a phase space region containing Kolmogorov-Arnold-Moser (KAM) islands generally obey a power law decay\cite{Lai1992}. 
Specifically, for an asteroid in the chaotic region of a restricted three-body problem (here the WD-Jupiter-asteroid configuration), the evolution of its energy follows a random walk and the number of escaping asteroids decays with the evolution time in a power law under the perturbation of Jupiter\cite{Sun1994,ZhouJL2000,ZhouJL2002}. 

Above dynamical evolution were studied with a bunch of N-body simulations.
The result of an example is plotted as brown line in Figure 1b, which matches well with the late stage of the observational best fit (blue line).
It worth remarking that such an example should represent a typical case.
The planet eccentricity set in this case ($e_{\rm p}=0.3$) is about the average eccentricity of exoplanets detected by radial velocity survey\cite{Wright2011},
and the total asteroid mass needed, i.e.,  $\sim$200 times the current mass of asteroid belt of solar system, is just several times the total mass of asteroids orbiting ${\zeta}$ Lep\cite{Chen2001}.
Beside the example case, we investigate more cases by vary various parameters, e.g.,  planet eccentricity, asteroid location and total asteroid mass in a large range.
For a large fraction of these cases, we still obtain reasonable fit to the bulk of data (See Supplementary Figure 7 in Methods).\\

It is worth pointing out that above simplified theoretical considerations may only catch the basic physics of the accretion picture and provide only the baseline of the accretion evolution.
In reality, the accretion history should be more complicated and many effects (i.e., magnetic field\cite{Farihi2017b}, dynamical instability of planets\cite{Debes&Sigurdsson2002, Veras2013}) might contribute more or less. These effects are expected to cause large variations on top of the baseline accretion, which may explain the relatively large scatters around the best fit as shown in Figure \ref{fig1}.

\begin{figure*}[!t]
\centering
\includegraphics[width=16cm]{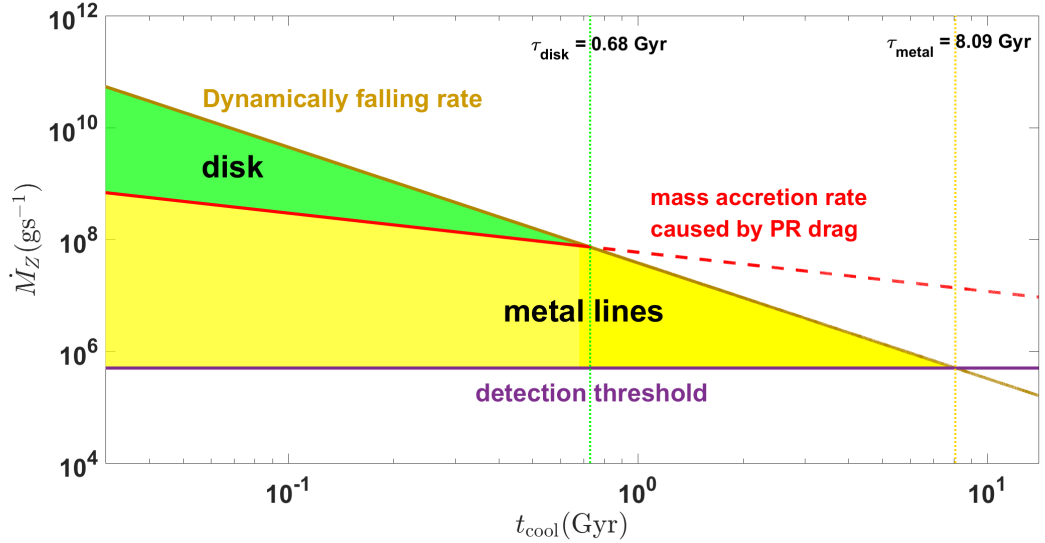}
\caption{{\bfseries A power law decay evolution scenario for polluted white dwarfs.} The brown line denotes the rate of mass falling into the Roche radius of the central WD derived from the best fit of observational data (Eqn. (\ref{dotMz}). The mass accretion rate caused by the PR drag is plotted as the solid red line, together with its extrapolation to older cooling ages as a dash red line  (i.e., Eqn. \ref{MPRt}). The horizontal purple line represents the detection threshold. The intersections of these lines provide estimates of the timescales of disk lifetime (green area) and metal line detection (yellow area). The vertical dotted lines denote the typical values of the timescales for dusty lifetime (green) and metal line detection (yellow).}
\label{fig2}
\end{figure*}

\subsection*{An evolution scenario}
\begin{figure*}[!t]
\centering
\caption{{\bfseries An evolution diagram for tidally disrupted asteroid model}}
\includegraphics[width=12cm]{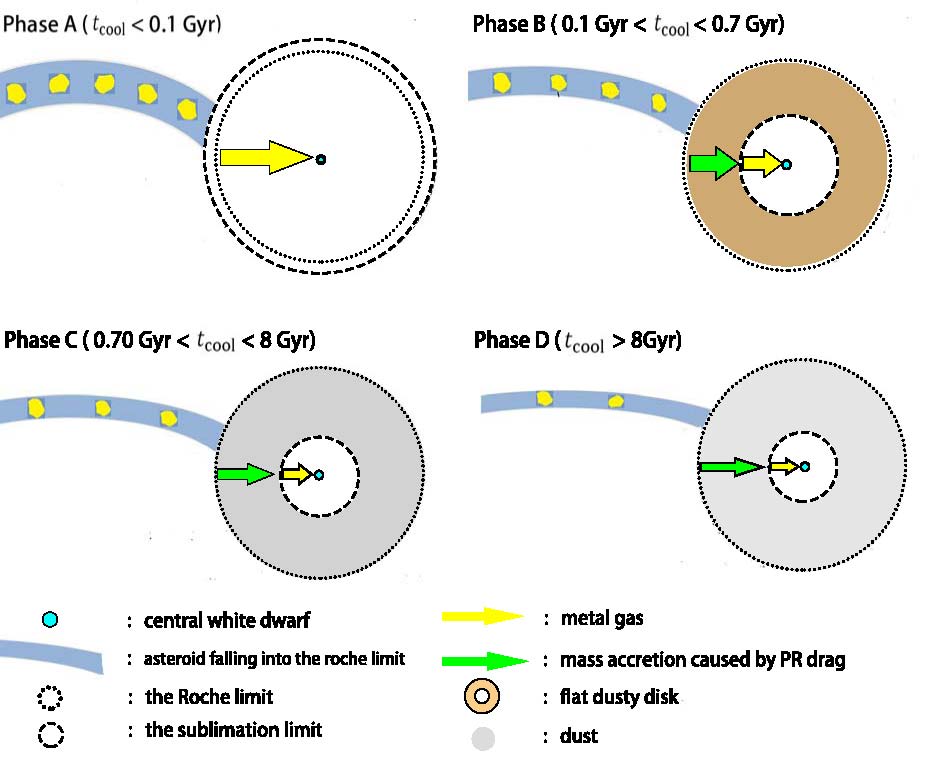}
\label{fig3}
\end{figure*}

The above match between observation and theory motivates us to outline the following evolution scenario, with the essence illustrated in Figure \ref{fig2} and Figure \ref{fig3}.

As shown in Figure \ref{fig2}, at the early stage, the rate of dynamically falling asteroids is larger than the maximum accretion rate driven by PR, i.e., ($\dot{M}_{\rm DF}>\dot{M}_{\rm PR}$), thus the dust can accumulate in the Roche  limit and form an opaque disk continually supplying accretion at a rate of $\dot{M}_{\rm PR}$\cite{Rafikov2011a}. At the late stage, when the $\dot{M}_{\rm DF}$ drops below $\dot{M}_{\rm PR}$, there is not enough material supply, thus the disk begins dissipating and becomes optically thinner, which in turn reduces the accretion rate\cite{Bochkarev2011} to keep balance with the income material. The transition time between these two stages therefore can be considered as the timescale of the disk phase, i.e., $\tau_{\rm disk}$.  Taking the turnoff time of the observed $\dot{M}_{\rm Z}$ in Figure 1b, $\tau_{\rm disk} = 0.68^{+0.25}_{-0.20}$ Gyr.

Interestingly, this is consistent with another observational fact shown in Figure 1b, namely none of disk-bearing DA WDs have cooling age greater than 0.95 Gyr. Nevertheless, one generally cannot expect to always detect disk for all WDs with $t_{\rm cool}<\tau_{\rm disk}$. 
Indeed, as shown in Figure 1b, a significant fraction ($\sim50\%$) of WDs  with cooling age less than $\sim$ 0.1 Gyr do not have detected disk. 
One possible reason could be that such young (thus hot) WDs had their dust disks significantly sublimated to be narrower and/or optically thinner, which are more difficult to be detected\cite{Bonsor2017}(see section 4.5 of Methods). Finally, the accretion rate $\dot{M}_{\rm Z}$ will drop to a degree that metal lines in WD atmospheres become too weak to be detected.
Here, we take the lowest measured mass accretion rate of our subsample with only DA WDs as the detection threshold, ${\dot{M}_{\rm dt}} = 10^{5.7} \rm gs^{-1}$. By equalizing $\dot{M_{\rm Z}}$ and $\dot{M}_{\rm dt}$, we obtain the critical timescale of detectable metal pollution:
$\tau_{\rm metal}=8.09^{+2.16}_{-1.65}$ Gyr, corresponding to a WD with $T_{\rm eff} \sim {\rm 4000K}$, which is consistent with WD 2251$-$070 ($T_{\rm eff}={\rm 4000K}$), the coolest metal-polluted WD observed by the Spitzer IRAC\cite{Farihi2009}. 

\begin{center}
\begin{table}[!h]
\centering
{\footnotesize
\label{}
\begin{tabular}{clclclc} \hline
   & Phase &  &  Infrared excess &     &  metal lines &  \\  \hline
   &  A (sublimation)        &     & weak/moderate      &    &  strong & \\
   &  B (dusty disk)            &     & moderate/strong       &    &  strong & \\
   &  C (Polution) &     & weak     &    &  moderate  &\\
   &  D (material depletion)           &     & weak     &    &   weak & \\\hline 
\end{tabular}}
\end{table}
\end{center}

In Figure \ref{fig3}, we divide the above evolutionary scenario into four observation phases and summarize the key observational features of each phase.   
In phase A ($t_{\rm cool}< \sim 0.1 {\rm Gyr}$), the central WD is so hot that the sublimation radius is comparable or even larger than the Roche  limit (See section 4.5 of Methods). 
After the asteroid being tidally disrupted into dust within the Roche  limit, most of the dust rapidly sublimates to gas and thus the infrared excess signal is weak/moderate. 
Nevertheless, the gas is still accreted by the WDs which produces strong metal lines in the observation.  
In phase B ($\sim 0.1{\rm Gyr} <t_{\rm cool}<0.7 {\rm Gyr}$),  the central WD is cool enough so that the sublimation radius is only about a few tenths of the Roche  limit. Sufficient dust accumulate and form a long-lived disk, supplying continuous dust accretion driven by PR, thus both the infrared excess and metal lines are strong.
In phase C ($0.7{\rm Gyr} <t_{\rm cool}< 8{\rm Gyr}$),  the rate of dynamically falling  asteroids drops and no longer sustains an opaque dust disk, thus the disk begins dissipating and becomes optically thinner with weak infrared excess. 
Although the dust accretion rate is reduced, it still pollutes the WDs by producing moderate metal lines.
In the final phase D ( $t_{\rm cool}>8{\rm Gyr}$), both the disk and the source material are almost exhausted, thus resulting infrared excess and metal lines are too weak to be detected.

In our scenario, the fraction of WDs with detectable disks to those with metal pollution is proportional to the fraction of their critical timescales. Therefore, the occurrence ratio of these two observational phenomena is predicted as
\begin{equation}
f_{\rm model} = \frac {{\tau}_{\rm disk}}{{\tau}_{\rm metal}}=8.41^{+4.44}_{-3.66}\%, 
\label{eq6}
\end{equation}
which is well consistent with the value derived from observations\cite{Koester2014,Barber2014,Zuckerman2003,Zuckerman2010}, $f_{\rm obs}= \frac {1\%-4\%}{25\%-50\%} \sim 2\%-16\%$. This is the first time that the occurrence ratio of these two observation phenomena has been explained quantitatively within one scenario.
Future survey with more detections of WDs with metal pollution and/or dusty disk will test and refine our scenario.

\subsection*{Data availability} 

\small The data that support the findings of this study are available upon request to the corresponding authors: Ji-Lin Zhou (zhoujl@nju.edu.cn) and Ji-Wei Xie (jwxie@nju.edu.cn).

\section*{Acknowledgements}
\small The authors thank the anonymous referees for useful suggestions that improved the manuscript. The authors thank Si-Yi Xu and Detlev Koester for useful discussions and suggestions. The authors also thank the Montreal White Dwarf Database (MWDD) for useful data and the evolutionary models of White dwarfs. Our research is supported by the National Natural Science Foundations of China (No.11333002, 11661161014, 11503009 and 11673011).  J.-W. X. also acknowledges  the Foundation for the Author of National Excellent Doctoral
Dissertation of People’s Republic of China (No. 10284201426) and the LAMOST Fellowship. 
The LAMOST Fellowship is supported by Special Funding for Advanced Users, budgeted and
administrated by Center for Astronomical Mega-Science, Chinese Academy
of Sciences (CAS).

\section*{Author contributions}
\small J.-L. Z. and D.-C. C. came up with the idea; D.-C. C., J.-W. X., M. Y. investigated the observation data and pertinent literatures; J.-W. X., D.-C. C and H. Z. conduct data analysis; D.-C. C. and Z.-Y. Y. conduct simulations and calculations; J.-L. Z., J.-W. X. and D.-C. C. wrote the paper; M. Y., H. Z., E.-S. L. and J.-Y. Y. participated in the revision of the paper.

\section*{Competing interests}
\small The authors declare no competing financial interests.

\section*{Methods}
\label{Methods}
{\bf \large \center Table of Contents\\}
\noindent
 {\bf (1) The samples.} (Supplementary Table 1, Supplementary Figure 1)\\ \\
 {\bf (2) The $\dot{M}_{\rm Z}$--$t_{\rm cool}$ relation derived from observation.} (Supplementary Table 2, Supplementary Figure 2-4)\\ \\
 {\bf (3) The $\dot{M}_{\rm Z}$--$t_{\rm cool}$ relation from theory.} (Supplementary Table 3)\\
       3.1 $\dot{M}_{\rm PR}$: mass accretion rate caused by PR drag. (Supplementary Figure 5) \\
       3.2 $\dot{M}_{\rm DF}$: dynamical falling rate delivered by the asteroids. (Supplementary Figure 6-7)\\ \\
 {\bf(4) Further discussions.} \\
       4.1 Observation bias. (Supplementary Figure 8-9)\\
       4.2 Conditions to escape disk detection\\
       4.3 Disk lifetime V.S. Disk phase time\\
       4.4 The influence of metal sinking timescale. (Supplementary Figure 10)\\ 
       4.5 The sublimation of asteroids. \\
       4.6 Other mechanisms for mass accretion. \\ 
       4.7 Previous studies on accretion rate $-$ cooling age relation.

\subsection*{1. The Sample}

We select our white dwarf (WD) sample based on the Montreal White Dwarf Database(MWDD) retrieved before May 24th, 2018.
First, we select all WD with reported atmospheric metal observation and effective temperature (corresponding to cooling age).
We then exclude WDs labeled as binary because the structure and evolution of circumbinary dusty disks and planetary systems may be different and more complex as compared to case in single star systems\cite{Farihi2017a}. 
For the majority of white dwarfs, calcium is the only atmospheric metal which could be detected in the optical at relatively cool stellar effective temperature and the calculations of mass accretion rates have been historically tied to calcium\cite{Farihi2016a}. 
Therefore we only choose the single WDs with measured atmospheric calcium abundance and effective temperature to make a uniform analysis across the (essentially) entire population of metal-polluted WDs.
We further exclude the 230 metal polluted white dwarf (all are non-DA) reported in Hollands et al. (2017)\cite{Hollands2017} because the derived uncertainties (about 30\%) of cooling ages of these 230 white dwarfs are systematically much larger (due to their abnormally large errors reported in their effect temperatures) than the others (about 10\%). 

Our sample consists of 47 single hydrogen-atmosphere (DA) and 799 helium-atmosphere (non-DA) WDs. 
These WDs are identified by two telescopes, 47 DA and 57 non-DA WDs identified by Spitzer IRAC in its first seven cycles\cite{Farihi2009,Farihi2010b,Xu2012,Girven2012,Bergfors2014}, 742 non-DA WDs identified by the Sloan Digital Sky Survey (SDSS) in Data Release (DR) 1, DR4, DR7, DR10, DR12\cite{Dufour2007,Koester2015,Limoges2015,Kepler2015,Kepler2016}.

With the assumption of steady accretion phase, the mass accretion rates are calculated as follows\cite{Farihi2016a}:
\begin{eqnarray}
\dot{M}_{\rm Z}=\frac{1}{\rm A_{\rm Ca}}\frac{\rm X_{\rm Ca}M_{\rm cvz}}{t_{\rm sink}(\rm Ca)}, 
\label{eqMz1}
\end{eqnarray}
  
where $t_{\rm sink}(\rm Ca)$ is the sinking timescale for calcium, $\rm A_{\rm Ca}$ represents the mass fraction of Calcium to the total accretion element, which is assumed as the bulk terrestrial mass fraction (0.016)\cite{Zuckerman2010,Farihi2016a}. $X_{\rm Ca}$ represents the mass ratio of atmospheric calcium to the convection zone, $M_{\rm cvz}$. The $t_{\rm sink}(\rm Ca)$, $M_{\rm cvz}$ and $t_{\rm cool}$ are functions of effective temperature $T_{\rm eff}$ and surface gravity $\log g$. 
Only a samall part of WDs in our sample have surface gravity and mass measurements. 
For those without such measurements, we assume them to be of a typical mass of $0.6M_{\bigodot}$ with $\log g = 8.0$\cite{Kepler2007}. 
Then, we obtain $t_{\rm sink}(\rm Ca)$, $M_{\rm cvz}$ and $t_{\rm cool}$ with the evolutionary models of WDs in MWWD\cite{Dufour2017}.
The mass accretion rate, $\dot{M}_{\rm Z}$ is:
\begin{eqnarray}
\dot{M}_{\rm Z}=6.25 \times 10^{7} \left(\frac{\rm A_{\rm Ca}}{0.016}\right)^{-1}\frac{\rm X_{\rm Ca}}{10^{-7}}\frac{M_{\rm cvz}/{t_{\rm sink}(\rm Ca)}}{10^{13}} {\rm gs^{-1}}, 
\label{eqMz2}
\end{eqnarray} 

The typical uncertainty of the observed calcium abundance from the Spitzer IRAC observations is 0.1 dex, causing a typical uncertainty (23\%) in $X_{\rm Ca}$\cite{Farihi2009}. 
While for SDSS observations, the typical uncertainty in the calcium abundance is 0.11 dex, causing a typical error (25\%) in $X_{\rm Ca}$. 
The ${M_{\rm cvz}/{t_{\rm sink}(\rm Ca)}}$ is a function of effective temperature $T_{\rm eff}$ and surface gravity $\log g$. There are 17975 single WDs with surface gravity measurement in the MWWD, whose effective temperatures range from 4000K to 23000K. Their surface gravity $\log g$ peak at 8.0 with a 1 $\sigma$ uncentainty of 0.33 (Supplementary Figure 1). With the evolutionary models of WDs in MWWD\cite{Dufour2017}, we obtain the uncertainty in ${M_{\rm cvz}/{t_{\rm sink}(\rm Ca)}}$ due to the uncertainty in $\log g$ and $T_{\rm eff}$ for each WD in our sample. By means of  error propagation, we obtain the uncertainty in $\dot{M}_{\rm Z}$ due to the uncertainties in $X_{\rm Ca}$ and ${M_{\rm cvz}/{t_{\rm sink}(\rm Ca)}}$. The typical uncertainty in $\dot{M}_{\rm Z}$ is 53\% for DA WDs and 32\% for non-DA WDs. Note, this uncertainty in $\dot{M}_{\rm Z}$ should be treated as a conservative value because we do not consider the dispersion in the elemental composition of accreted material (i.e., $\rm A_{\rm Ca}$) and this could induce extra uncertainty in $\dot{M}_{\rm Z}$.  The derived accretion mass rates and cooling ages of WDs in our samples are listed in the Supplementary Table 1. \\
\subsection*{2. The $\dot{M}_{\rm Z}$--$t_{\rm cool}$ relation derived from observation}   In this section,  we fit the relation between $\dot{M}_{\rm Z}$ and $t_{\rm cool}$ with three models: constant, single power law and broken power law. The Formula forms of the three models are:

\begin{eqnarray}
\log\left(\frac{\dot{M}_{\rm Z}}{\rm gs^{-1}}\right) = {\rm B_0}, 
\label{eqmodel1}
\end{eqnarray}

\begin{eqnarray}
\log\left(\frac{\dot{M}_{\rm Z}}{\rm gs^{-1}}\right) = {\rm A}\times \log\left(\frac{t_{\rm cool}}{\rm Gyr}\right)+ {\rm B}, 
\label{eqmodel2} 
\end{eqnarray}

\begin{eqnarray}
\log\left(\frac{\dot{M}_{\rm Z}}{\rm gs^{-1}}\right) =
\begin{cases}
{\rm A_1}\times \log\left(\frac{t_{\rm cool}}{\rm Gyr}\right)+ {\rm B_1},& \text{$t_{\rm cool}<t_{12}$}\\
{\rm A_2}\times \log\left(\frac{t_{\rm cool}}{\rm Gyr}\right)+ {\rm B_2},& \text{$t_{\rm cool}\ge t_{12}$}
\end{cases}. 
\label{eqmodel3}
\end{eqnarray}

For each model, we fit the relationship between $\dot{M}_{\rm Z}$ and $t_{\rm cool}$ of WDs in the given sample with the Levenberg-Marquardt algorithm (LMA). 
In order to avoid being trapped in some local best fit, we randomly generate 1000 set of initial starting guesses of the fitting parameters before LMA fitting. 
We then select the one with lowest residual sum of squares (RSS) from the 1000 local best-fits as the global best fit.
In order to compare the global best fits of different models,  we calculate the Akaike information criterion (AIC) for each model. In terms of RSS,  the AIC is calculated as:\cite{Cavanaugh1997}:
\begin{eqnarray}
AIC = 2k+n\ln(RSS) , 
\label{AIC}
\end{eqnarray}

where n is the number of data points, k is the number of model parameters. 
 To obtain the uncertainty of each fitting parameters, we assume that the observed mass accretion rates obey the Gaussian distribution $N(\dot{M}_{\rm Z}, {\sigma}^2)$, where $\dot{M}_{\rm Z}$ and $\sigma$ are the observed accretion rate and corresponding uncertaity. Then we resample the observed accretion rates from these Gaussian distributions and refit the data. We repeat this procedure 1000 times and obtain 1000 sets of best fits and their respective AIC scores.
The uncertainty (one-sigma interval) of each parameter is set as the range of $50\pm34.1$ percentiles of the 1000 sets of best fits.  

We apply above data fitting procedure to three cases: the whole sample, subsample with only DA WDs and subsample with only non-DA WDs respectively. 
The results are shown in Supplementary Figure 2-4 with the fit parameters and the AIC values summarized in Supplementary Table 2. 
For the 1000 sets of resample data of each case, the constant model always has much larger AIC values (AIC difference $\Delta \rm AIC>10$) and thus can be confidently ruled out in all the cases.
Although a broken power law is preferred if using the whole sample data ($\Delta \rm AIC \sim 77$), such a result is likely to be unreliable. 
As shown in Supplementary Figure 2,  for the first 0.1 Gyr  ($t_{\rm cool}<0.1$ Gyr),  the data is dominated by DA WDs with $\dot{M}_{\rm Z}$ around $10^9\rm gs^{-1}$, while for the subsequent 0.1 Gyr  (0.1 Gyr $<t_{\rm cool}<$ 0.2 Gyr),  it is dominated by non-DA WDs with $\dot{M}_{\rm Z}$ around $10^{10}\rm gs^{-1}$.
Such systematical differences between DA and non-DA WD should be the cause of the increasing part of the broken law. 
Indeed, if we only take the subsample of non-DA WD, we see (Supplementary Figure 3) the broken power law model essentially reduces to the single power law model, with the increasing part caused by only a few outliers.  

For the subsample of DA WDs, a broken power law model is preferred against a single power law by an AIC difference of $\Delta \rm AIC \sim 2$ (Supplementary Figure 4).  Among the 1000 sets of the resampled data, a broken power law model is preferred with a smaller AIC score for 973 sets, corresponding to a confidence level of 97.3\%.   
Although the significance (about 2 sigma) is not very strong, the best fit of the broken power law is very encouraging as its first part match almost exactly with theoretical expectation of dust accretion driven by the Poynting-Robertson (PR) effect (Figure 1b).

In any case, a power law decay model fits the data well at late stage (i.e., $t_{\rm cool}>1$ Gyr) regarding less of whether DA or non-DA WDs.
Nevertheless, we still prefer to use the subsample of DA WDs for the following reasons.
(1) DA WDs cover a larger range of cooling age (especially in the early end) and better to study the evolution of accretion rate. (2) The $\dot{M}_{\rm Z}$ derived for non-DA only reflects the historical average value, which could be severely some rare outliers with extremely lager accretion rate, e.g., runaway accretion\cite{Rafikov2011b}. (3) The $\dot{M}_{\rm Z}$ derived for DA WDs reflects the on-going accretion rate (outliers are likely to be filtered out), thus better for tracing the baseline of accretion evolution.  \\

\subsection*{3. The $\dot{M}_{\rm Z}$--$t_{\rm cool}$ relation from theory} 

In the tidally disrupted asteroid model\cite{Jura2003,Farihi2016a}. Asteroids beyond several AU in the remnant planetary system may pass through the Roche limit of the central WD in highly eccentric orbits as they are continuously perturbed and be torn apart to rubbles by gravitational tides. Due to mutual collisions and dynamical relaxing, a flat circumstellar disk forms with dust and will be detected as an infrared excess under stellar irradiation\cite{Debes&Sigurdsson2002,Veras2014}. In the subsequent evolution, closely orbiting dust will drop onto the stellar surface mainly due to the Poynting--Robertson (PR) drag, which are detected as metal lines\cite{Koester2014,Farihi2016a}. Therefore the mass accretion rates calculated from the metal lines are determined by the accretion rate caused by the PR drag $\dot{M}_{\rm PR}$ and the dynamical infall rate of asteroid into the Roche limit $\dot{M}_{\rm DF}$.

{\bfseries \noindent 3.1 $\dot{M}_{\rm PR}$: mass accretion rate caused by PR drag.}  Assuming a steady accretion phase for an opaque disk with a narrow optical transition zone, the mass accretion rate caused by PR drag is\cite{Rafikov2011a}:
\begin{eqnarray}
\dot{M}_{\rm PR}=\frac{32\phi _{r}}{3}\sigma(\frac{R_{*}T_{\rm eff}T_{s}}{c})^2  ,  
\label{eqS7}
\end{eqnarray}
where $R_{*}$ and $T_{\rm eff}$ are the radius and stellar effective temperature of the center WD, $T_{s}$ is the sublimation temperature of asteroids. $\sigma$ and $c$ are the Stefan-Boltzmann constant and velocity of light. Taking $\phi _{r}=1$ and $T_{\rm S}=1200 \rm K$\cite{Xu2012}, we obtain the relationship between $\dot{M}_{\rm PR}$ and $T_{\rm eff}$:
\begin{eqnarray}
\dot{M}_{\rm PR}=1.8\times 10^{8}(\frac{T_{\rm eff}}{\rm 15000K})^2 \frac{M}{0.6M_{\bigodot}} \left(\frac{g}{10^{8}\rm cms^{-2}}\right)^{-1}  {\rm gs^{-1}}. 
\label{eqS8}
\end{eqnarray}
To obtain the relationship between $\dot{M}_{\rm PR}$ and $t_{\rm cool}$, we substitute $T_{\rm eff}$ with $t_{\rm cool}$ using the following form. 
When a star evolve into an WD, the residual ion thermal energy is the only significiant source of radiation energy.  For WDs with luminosities less than $10^{-0.5}L_{\bigodot}$ (solar luminosity) ($T_{\rm eff} \lesssim 38000\rm K$), the radiation would be dominated by photon emission and their cooling ages are\cite{Shapiro1983}
\begin{eqnarray}
t_{\rm cool}=\frac{3}{5}\frac{{\rm k}T_{*}M}{{\rm A}m_uL}, 
\label{eqS9}
\end{eqnarray}
\begin{eqnarray}
T_{\rm *}=\left(\frac{L}{1.2\times 10^{6}{\rm ergs^{-1}}}\right)^{\frac{2}{7}}\left(\frac{M}{0.6M_{\bigodot}}\right)^{-\frac{2}{7}}, 
\label{eqT_*}
\end{eqnarray}
where k is the Boltzmann constant and ${m_u}$ is the atomic mass unit. $T_{*}, M, L$, A are the interoir temperature, mass, luminosity and average baryon number, respectively.  The cooling of WD would  become more rapidly due to lattice vibrations of the ions only for sufficiently low L ($\lesssim 10^{-4.5}L_{\bigodot}, T_{\rm eff} \lesssim 3800\rm K$). 

The $T_{\rm eff}$ of WDs in our samples, ranges from 23000K to 4000K. Therefore, Eqn. (\ref{eqS9}) are appliable for all WDs in our sample. Using $L= 4\pi R^{2}\sigma {T_{\rm eff}}^4$ and $g=\frac{GM}{R^2}$, we get
\begin{eqnarray}
\log\left(\frac {t_{\rm cool}}{\rm Gyr}\right)=11.24-\frac{20}{7}\log\left(\frac{T_{\rm eff}}{\rm K}\right)+\frac{5}{7}\left(\log g-8.0\right).
\label{eqS10}
\end{eqnarray}

A KS test is done between $t_{\rm cool}$ from Eqn. (\ref{eqS10}) and the data in MWDD and the calculated confidence level is lager than $99.5\%$. Supplementary Figure 5 depicts the comparison and the KS test result between $t_{\rm cool} (\rm Gyr)$ from Eqn. (\ref{eqS10}) and the data in the MWDD. By combining Eqn. (\ref{eqS8}) and Eqn. (\ref{eqS10}), we obtain: 
\begin{eqnarray}
\log\left(\frac {\dot{M}_{\rm PR}}{\rm gs^{-1}}\right)=-0.7\times \log\left(\frac{t_{\rm cool}}{\rm Gyr}\right)+7.77+\log \left(\frac{M}{0.6M_{\bigodot}}\right)+0.5\left(\log g-8.0\right). 
\label{eqPRr}
\end{eqnarray}
The surface gravity $\log g$ stronly peaked at 8.0 and the uncentainty is 0.33, which would cause that the masses of WDs are distributed mainly between 0.4 $M_{\bigodot}$ and 0.8 $M_{\bigodot}$, with a strong peak at $0.6M_{\bigodot}$\cite{Livio2005,Kepler2007}. Here we set 0.2 $M_{\bigodot}$ as the typical error for WD mass. Taking $M_*=0.6M_{\bigodot}$ and $\log g=8.0$ with their uncertaities, by means of error propagation, we obtain Eqn. (\ref{MPRt}), which describes the relationship between $\dot{M}_{\rm PR}$ and $t_{\rm cool}$:  
\begin{eqnarray}
\log\left(\frac {\dot{M}_{\rm PR}}{\rm gs^{-1}}\right)=-0.70\times \log\left(\frac{t_{\rm cool}}{\rm Gyr}\right)+7.77(\pm 0.31).
\end{eqnarray}

{\bfseries \noindent 3.2 $\dot{M}_{\rm DF}$: dynamical falling rate delivered by the asteroids. \\} To find the relationship between the $\dot{M}_{\rm DF}$ and $t_{\rm cool}$ , we use a simplified solar system remnant as a fiducial model, which consists of a central WD, an asteroid belt and a Jupiter-mass planet. 
The mass of the central WD was 0.69$M_{\bigodot}$ (same mass as G29-38, the first detected WD with an infrared excess)\cite{Jura2003} and the Roche limit is 1.2$R_{\bigodot}$. 
For a typical WD progenitor ($\thicksim 2M_{\bigodot}$ A-type star)\cite{Catalán2008,Koester2014}, its envelope would expand to $\thicksim 2 {\rm AU}$ in the red-giant stage\cite{Veras2016}. 
The planets in the inner region would be swallowed, while the orbits of outer planets would expand by a factor of $\thicksim 3$ due to stellar mass loss\cite{Mustill2012}. 
Thus, the inner edge of the asteroid belt should be beyond 6 AU and  the semi-major axes of the Jovian planet is three times that of the current Jupiter, i.e., $a_{\rm p} = 15.6$ AU. 
The orbital eccentricity of the planet is set as a free parameter range from 0.1 to 0.5. 
Larger eccentricities are omitted to avoid the planet pass through the whole asteroid belt.

Asteroids in the chaotic zone would undergo chaotic motion, which could be ejected from the system through close encounter with the planet or may enter the Roche limit of the central WDs\cite{Frewen2014}. 
Therefore, we focus on the asteroids in the chaotic zone  of the Jovian planet.
The orbits of small bodies in the restricted circular three-body problem were chaotic when\cite{Duncan1989}
\begin{eqnarray}
\varepsilon = \frac {|a-a_{\rm p}|}{a_{\rm p}} \leqslant 1.5{(\frac {m}{M_{*}})}^{\frac {2}{7}}, 
\label{eqS11}
\end{eqnarray}
where $a$, $a_{\rm p}$ are the semi-major axes of the small body and the planet, $m$, $M_{*}$ were the mass of the planet and central star, respectively. For the elliptical restricted three body problem, the inner edge $a_{\rm in}$  of chaotic zone is\cite{Frewen2014}:
\begin{eqnarray}
a_{\rm in} = a_{\rm p}(1-\varepsilon _{\rm in}), 
\label{eqS12}
\end{eqnarray}
\begin{eqnarray}
\varepsilon _{\rm in} = \frac {a_{\rm p}-a_{\rm in}}{a_{\rm p}} = 1.5{(\frac {m}{M_{*}})}^{\frac {2}{7}} + e_{\rm p}(1-1.5{(\frac {m}{M_{*}})}^{\frac {2}{7}}).
\label{eqS13}
\end{eqnarray}

In each set of simulation, we put 20,000 asteroids (treated as test particles) initially in the chaotic zone with a width of 1 AU in semi-major axis.
The initial orbits of asteroids are coplanar and circular with other orbital elements randomly drawn from 0 to 360 degree. 
We use N-body simulation to follow their orbital evolutions. 
The simulations are carried out by using the MECURY package\cite{Chambers1999} with the Bulirsch-Stoer (BS) algorithm\cite{Press1992}, which was specifically designed for accurate calculations of high-eccentricity orbital evolutions and close encounters.  

During the simulations, the asteroids were removed from the system when they (1) fell into the Roche limit of the central WDs, i.e., distance to the star $<0.006$ AU; (2) ejected from the system, i.e., distance to the star $>100$ AU; (3) collided with the Jovian planet, i.e., distance to the planet less than the radius of Jupiter. Upon completion of the simulations, the returned information consist of the removal mechanisms and the lifetimes of each asteroid.

In Supplementary Figure 6, we plot $P_{\rm F}$ (the ratio of the number of particles falling into the Roche limit per year) as a function of time (e.g., $t_{\rm cool}$) for several typical cases.
We see that the evolution of $P_{\rm F}$ can be well fit with a power law decay function, i.e., 
\begin{eqnarray}
\log\left(P_{\rm F}\right)=\alpha \times \log\left(\frac{t_{\rm cool}}{\rm Gyr}\right)+\beta, 
\label{eqS14}
\end{eqnarray}
where $\alpha$ and $\beta$ are two coefficients which depend on the model parameters set in the simulation as shown in Supplementary Table 3. 
Generally, larger planet eccentricity and closer distance to between the planet and the asteroid belt result steeper decay (smaller $\alpha$ and $\beta$). 
Limited by the computation cost, most of simulations end at $t_{\rm cool}=0.1$ Gyr. 
Only one simulation (green dot in Supplementary Figure 6) initially with 100, 000 asteroids ran for $t_{\rm cool}=1$ Gyr.
And the result demonstrates that the power law decay can be extended to longer timescale on order of Gyr.

The rate of  dynamically falling asteroids, $\dot{M}_{\rm DF}$, is calculated as a scale of $P_{\rm F}$, i.e., 
\begin{eqnarray}
\dot{M}_{\rm DF} = P_{\rm F}\frac{M_{\rm belt}}{M_0}\frac{3.6\times 10^{24}}{3.15 \times 10^{7}} {\rm gs^{-1}} , 
\end{eqnarray}
where $M_{\rm belt}$ is the mass of the asteroid belt initially around  the WD, $M_{0}=3.6\times 10^{24}$ the mass of the main asteroid belt of the solar system\cite{Krasinsky2002} and $3.15 \times 10^{7}$ the number of seconds per year.
In Supplementary Figure 7, we plot $\dot{M}_{\rm DF}$ as a function of $t_{\rm cool}$ on top of the observation data from the subsample of DA WDs. 
As can be seen, the observation data cover a large range of the model parameters.
In other words, one dose not have to fine tune the model parameters (e.g., the planet eccentricity) to fit the observation data.
Nevertheless, we can still obtain some loose constraint on the initial mass of the asteroid belt, i.e., $10 M_0<M_{\rm belt}<5000 M_0$, which is consistent with the estimates of previous studies\cite{Frewen2014,Debes2012}.
It is worth noting that asteroids in the chaotic zone could be significantly depleted during the main sequence evolution\cite{Morrison2015}.  Nevertheless,  the chaotic zone could be widened during the post main sequence evolution by a factor of $\sim 1.5$ to include more source  asteroids\cite{Frewen2014,Mustill2014}. Therefore, our above mass constraint on the initial mass of asteroids is generally underestimated by a factor of $\sim3$.      
\subsection*{4. Further discussions}
{\bfseries \noindent 4.1 Observation bias. \\}
Because the ionization of CaII to CaIII increases with the effective temperature, the detection limit for Ca abundance increases with the effective temperatures\cite{Koester2006}.  Since the derivation of accretion rate ($\dot{M}_{\rm Z}$) is based on Ca detection,  thus there is bias against low abundance (low accretion rate) for hot (young) WDs. Such a bias could potentially result in a decay trend in the observed accretion rate.   
Following Koester \& Wilken (2006)\cite{Koester2006}, we obtained the observation limit for Ca II K line  with an equivalent width 15 m$\mathring{\rm A}$ for DA WDs. With Equation S2, we transfer the detection limit of Ca II K line to the detection limit of $\dot{M}_{\rm Z}$ for a typical WD ($0.6M_{\bigodot}$ and $\log g=8.0$).  Due to the much larger opacity of hydrogen than helium, metal lines in non-DA WDs are much stronger thus much easier to be detected. Therefore the detection limit should be even lower for non-DA WDs\cite{Zuckerman2003,Dupius1993}.  

To better see the effect of this bias, we plot $\dot{M}_{\rm Z}$ as a function of $T_{\rm eff}$ and $t_{\rm cool}$ in Supplementary Figure 8 and Supplementary Figure 9.
The orange line shows the detection limit caused by the above observational bias.
We argue that the observed accretion rate decay is essentially not due to observational bias given the following reasons. 
First, the ionization of CaII is not significant for cool WDs with effective temperature less than 13,000K, correspond to $t_{cool}> 0.3 \rm Gyr$, and thus, the observational bias should be negligible (detection limit line becomes much flatter).  
In contrast, the accretion decay shown in the data is most significant in the cooler end.
Second, the accretion decay can also be clearly seen from the upper envelop of the data. \\

{\bfseries \noindent 4.2  Conditions to escape disk detection. \\}
We discuss the conditions for disk to escape detection.
Observationally, Spitzer and WISE are likely to miss to detect a disk in the following two situations\cite{Bonsor2017}, (1) an opaque dust disk but narrower than 0.04$r_{\rm in}$ (where $r_{\rm in}$ is the disk inner edge) and (2) the an optically thin disk.
The first situation may appear at the beginning, i.e., phase A as shown in Figure 3.
Metal gas dominates the region within Roche  limit due to sublimation, and the dust disk, if any, is very narrow and thus escapes detection.
The second situation may appear at late stages, e.g., phase C as shown in Figure 3.
The dust accretion rate is reduced by 1-2 orders of magnitude as compared to the accretion rate (Eqn. 2) of an opaque disk driven by PR.
Nevertheless such a lower accretion rate could be still driven by PR but for an optically thin disk\cite{Bochkarev2011}, which can escape detection. \\

{\bfseries \noindent 4.3  Disk lifetime V.S. Disk phase time. \\}
It worth pointing out that disk lifetime and disk phase time are defined differently.
Disk lifetime ($t_{\rm disk}$) usually refers to the lifetime of a disk that results from an individual tidally disrupted event.
For an asteroid with radius of 100 km and mass about $10^{22}$ g, assuming accretion rate of $10^8 \rm gs^{-1}$, the disk lifetime is about several Myr. For an asteroid with smaller mass ($<10^{22} \rm gs^{-1}$), the circumstellar disk would be opatically thin and characteristic timescale would be $10^{5}-10^{6}$ year\cite{Bochkarev2011}.
While in this paper, disk phase time ($\tau_{\rm disk}$) refers to the timespan of an evolutional phase in which the WDs is orbited by an {\it{opaque dusty}} disk most of the time.
There could be some disk-free intervals during the disk phase, but they should be short and rare.\\

{\bfseries \noindent 4.4 The influence of metal sinking timescale.\\}
If a WD is observed to have metal pollution, the probability that it is caused by a transient phase of historical accretion not by on-going accretion, can be estimated as the ratio of sinking timescale and the disk lifetime, i.e., $p_{\rm tran}\sim t_{\rm sink}/t_{\rm disk}$\cite{Farihi2016a}. 
According to the evolutionary models of MWWD\cite{Dufour2017}, we calculate the sinking timescale of calcium for a typical white dwarf ($0.6M_{\bigodot}$, $\log g = 8.0$)\cite{Kepler2007}.
Supplementary Figure 10 depicts the sinking timescales as a function of cooling age for different stellar spectral type.
Considering the typical disk lifetime\cite{Bochkarev2011}, $t_{\rm disk}=10^5-10^6$ yr, the transient probability, $p_{\rm tran}$, will approach unity for most non-DA (DB) WDs and old ($t_{\rm cool}>$ 2 Gyr) DA WDs, which causes two important results. 
First, as mentioned at the end of section 2, the observed accretion rate of non-DA WDs is only a historical average value, which could be dominated by outliers due to some rare accretion burst. 
From this angle of view, DA WDs with shorter $t_{\rm sink}$, are better tracers of accretion evolution.
Second,  old DA WDs with $t_{\rm cool}>$ 2 Gyr maybe currently disk-free though they are currently found to have metal pollution. 
The metal pollution could be caused by some historical accretion of metals, which are still stored in the atmosphere of WD since the metal sinking timescale is long.   \\

{\bfseries \noindent 4.5 The sublimation of asteroids. \\}  The sublimation radius, $R_{s}$ where the effective temperature equals the sublimation temperature of asteroids, $T_{s}$ is\cite{Rafikov2011a}:
\begin{eqnarray}
R_{s} = \frac{R_*}{2} \left(\frac{T_{\rm eff}}{T_{s}}\right)^2 \approx 22R_{*}\left(\frac{T_{\rm eff}}{10000\rm K}\right)^2 \left(\frac{T_s}{1500 \rm K}\right)^{-2}, 
\label{eqS16}
\end{eqnarray}
where $R_{*}$ and $T_{\rm eff}$ are the radius and effective temperature of WDs. The sublimation temperature of asteroids $T_{s}$ is usually 1200K to 1500K\cite{Rafikov2011a, Xu2012, Bonsor2017}.
For a typical white dwarf ($0.6M_{\bigodot}, \log g =8.0$), and the Roche  limit is about $0.8-1.2 R_{\bigodot}$\cite{Farihi2016a}.  
Equating the Roche  limit with the sublimation radius,  we obtain $T_{\rm eff} \sim 17000 (15000-23000)\rm K$, corresponding to the cooling age of $\sim 0.1 (0.04-0.21)$ Gyr, which sets the typical timescale of phase A as shown in the Figure 3 of the main text.
For very young ($t_{\rm cool}< \sim 0.1$ Gyr) and thus very hot WDs, $R_{s}$ could exceed the Roche  limit of WDs.  
Therefore, after the asteroid being tidally disrupted into dust within the Roche  limit, most of the dust rapidly sublimates to gas. 
The infrared excess would be weak and escape detection,
while the mass accretion could be supplied by pure gas accretion\cite{Bonsor2017}.  \\
    
{\bfseries \noindent 4.6 Other mechanisms for mass accretion. \\} 
Beside PR drag, the accretion of circumsteller dusty disks around WDs is influenced by the gas within the dust disk, the Yarkovsky force and stellar magnetospheres\cite{Rafikov2011a,Farihi2016a,Farihi2017b}. 
The sublimation of dust in the disk within the Roche limit would lead to the existence of gas\cite{Jura2008,Gänsicke2006,Gänsicke2007,Gänsicke2008,Melis2012}. The additional viscosity due to the presence of gas would enhance the accretion rate and potentially lead to runaway accretion\cite{Rafikov2011b,Metzger2012} with a rate of $10^{10}-10^{11} \rm gs^{-1}$. There are a number of non-DA white dwarfs with such high accretion rates. However, no such high accretion rate has been observed in DA WDs, which strongly suggests the short lifetimes and small detection possibility of such phase\cite{Koester2014}. 

Due to the thermal inertia of spinning dust, the reemission of absorbed stellar energy would be different from the radial in the direction. This will lead to the Yarkovsky force and causes outward orbital migration\cite{Bottke2006}. However the the Yarkovsky force plays an unimportant role in the evolution of disks comparing with the PR drag\cite{Rafikov2011a}. The magnetic interactions could strongly influence the dynamics of dust particles with size smaller than 1 $\rm \mu m $ for a stellar magnetic field of a few tens of kG\cite{Farihi2017b,Farihi2018}. Neverthless, for the bulk of the disk mass, which mainly consists of particles with larger sizes ranges from $\thicksim \rm 0.03-30 cm$\cite{Rafikov2012}, the magnetic interactions are essentially negligible\cite{Farihi2017b}. 

For a circumstellar disk produced by collisional cascades inside the Roche limit of WD, adding asteroids with radius $r_{\rm 0}$ into the disk with a rate $\dot{M_{\rm 0}}$, the accretion is steady and the mass accretion rate into the WD atmosphere is roughly $\dot{M_{\rm 0}}$ only when $r_{\rm 0}$ is smaller than $\thicksim \rm 10 km$\cite{Kenyon2017a,Kenyon2017b}. Assuming an asteroid belt with mass distribution proportional to $M^{-2}$ and particle size from $\rm 0.1-100 km$\cite{Bergfors2014,Kenyon2017a}, which is similar to main asteroid belt of the solar system, the majority of asteroids destroyed within the Roche limt are small\cite{Bergfors2014} and their accretion would be steady. Besides, Furthermore, this work is valid for prolate ellipsoids which are near the Roche limit of WDs and with aspect ratios of roughly $2:1:1$\cite{Kenyon2017a}. While the disruption timescale for spherical asteroids with $r=\rm 1-100 km$ within the Roche limit is days to years\cite{Richardson2005,Veras2017}. Therefore, even if the radius of spherical asteroids is larger than 10 km. they could be tidally disrupted to small rubbles ($r<\rm 1 km $) and the accretion would become steady.

{\bfseries \noindent 4.7 Previous studies on accretion rate $-$ cooling age relation. \\}  
There are a few other studies on the trend between the accretion rate and cooling age of metal-polluted WDs before. In  Koester \& Kepler (2015), 75 metal-polluted non-DA WDs are identified and their calcium accretion rates decrease strongly with decreasing effective temperature (cooling age)\cite{Koester2015}. Similar decay trend has been discovered in Kepler et al. 2015 with 397 non-DA samples from SDSS DR 10\cite{Kepler2015} and Kepler et al. 2016\cite{Kepler2016} with 236 non-DA samples from SDSS DR 12. 

In the literature Wyatt et al. (2014)\cite{Wyatt2014}, they considered a sample comprised of DA samples and a small non-DA samples and did not found a sufficient evidence for a decay trend. The DA samples contains 38 DA WDs with detection of calciums, 496 WDs with upper limits of Caliums from a Keck survey\cite{Zuckerman2003} and a SPY survey. The small non-DA samples consists of 12 non-DA WDs with detection of calciums and 18 upper limit from the Table 1 of Zuckerman et al. 2010\cite{Zuckerman2003}. They did not find a sufficient evidence for a decay trend with age. The possible reasons for that could be the following. (1) Most data in their samples only have upper limit.  (2) Their sample of WD with detection of Calium are small and young. For example, all non-DA WDs have cooling age less than 0.3 Gyr and only 2 DA WDs with age larger than 2 Gyr. While the decay shown in our work is most significant for WDs with cooling ages larger than 1 Gyr.

In the  Koester et al. (2014)\cite{Koester2014}, they did not found significant trend either. As can be seen in Fig. 8 of Koester et al. (2014), the decay trend is ruined mainly by including those very hot and young WDs (with $T_{\rm eff}>18,000 \rm K$). As we discussed in section 4.5 of the Method, such hot WDs will have its sublimation radius larger than the Roche radius, thus the disk is likely to be dominated by gas. The gas accretion could be affected by stellar magnetic field, and could be highly unstable and result in variable accretion flux\cite{Farihi2018}. This may explain the large range of accretion rates of those very young WDs as shown in Fig.8 of Koester et al. (2014). Besides, the accretion rates of these hot DAZs are calculated based on the Si abundance, while the 38 colder DAZ based on the Ca abundance. Thus, the discrepancy could be caused by some systematical difference between mass accretion from different elements. If only considering cooling age larger than 0.2 Gyr in their samples, a decay trend of mass accretion rate is generally seen.

\bibliographystyle{plain}

\begin{figure}[!h]
\centerline {\includegraphics[width=0.9\textwidth]{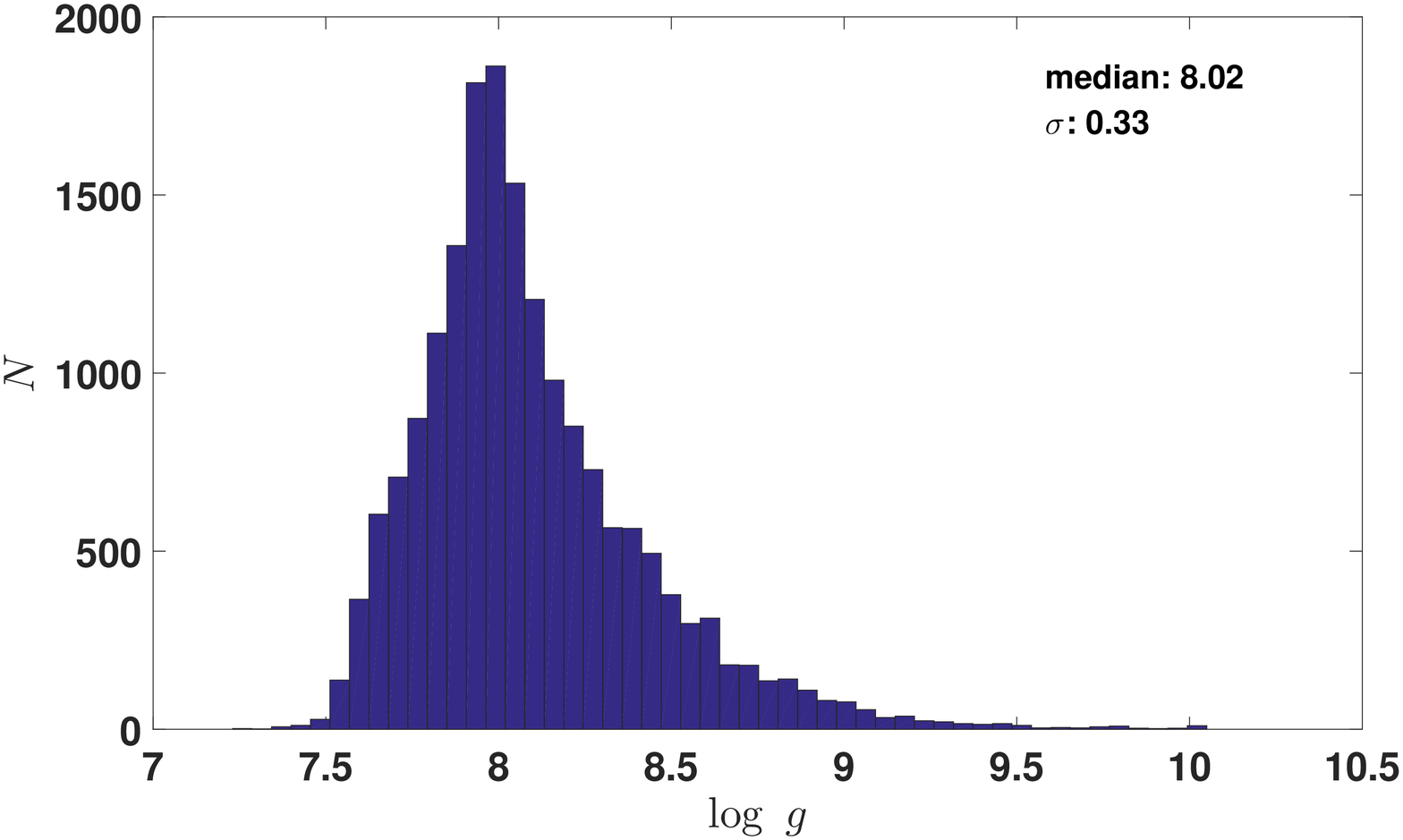}}
{Supplementary Figure 1: The distribution of measured surface gravity $\log g$ of single WDs with $T_{\rm eff}$ from 4000K to 23000K in MWWD.}
\label{flgg}
\end{figure}

\begin{figure}[!h]
\centerline {\includegraphics[width=0.9\textwidth]{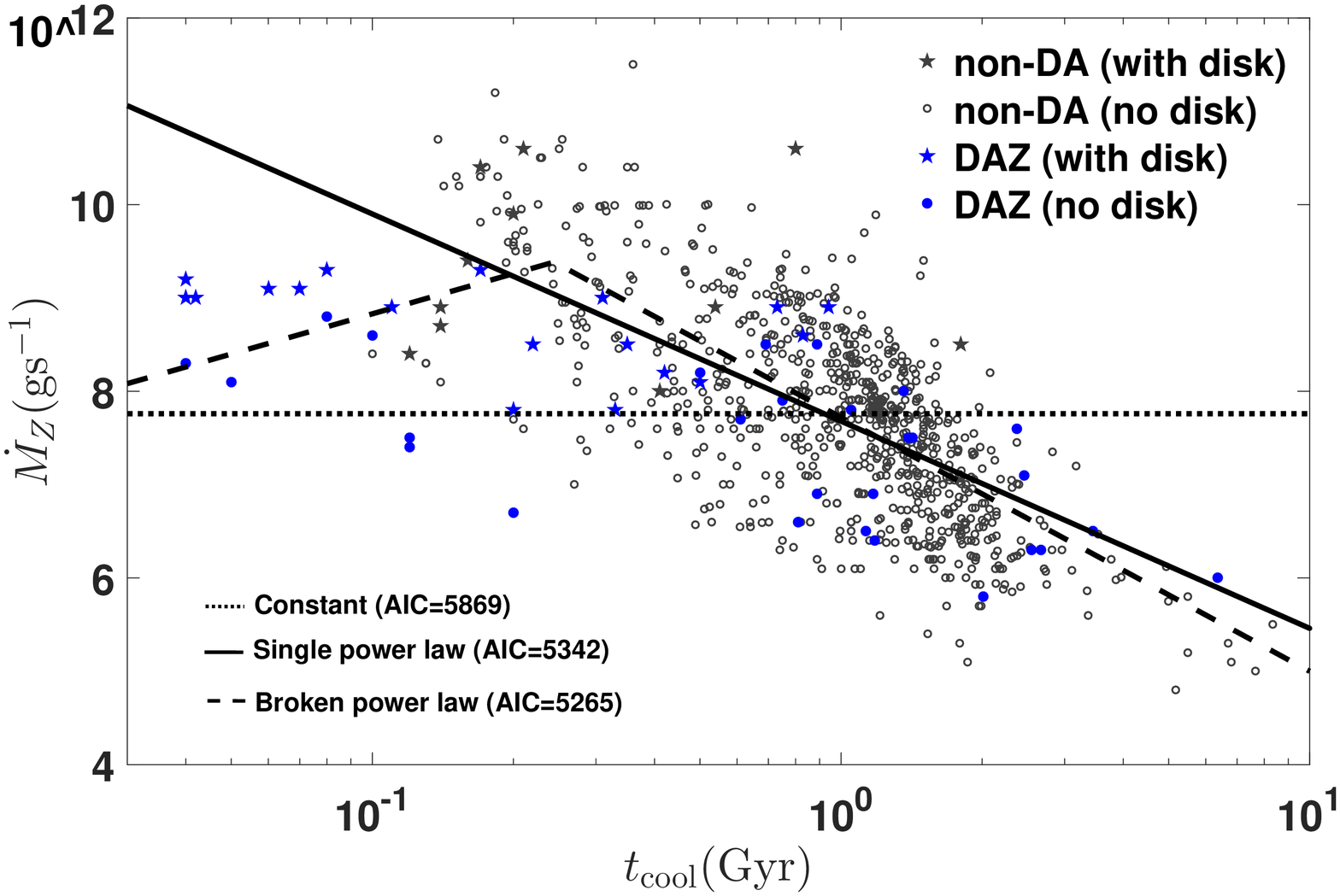}}
{Supplementary Figure 2: The mass accretion rate vs. the cooling ages of all 846 WDs. The three black lines donote the best fitting of each model for all WDs.}
\label{all}
\end{figure}

\begin{figure}[!h]
\centerline {\includegraphics[width=0.9\textwidth]{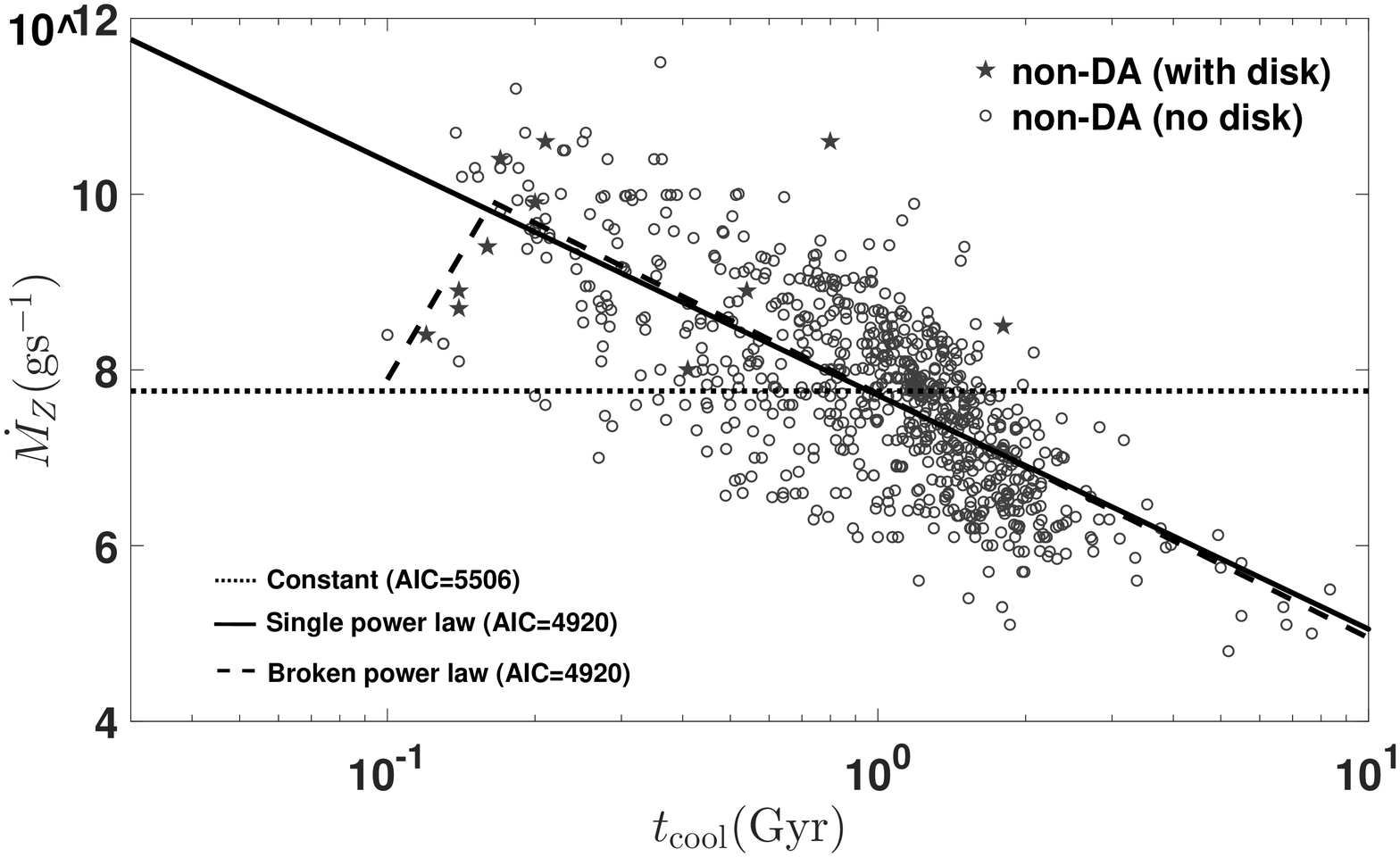}}
{Supplementary Figure 3: The mass accretion rate vs. the cooling ages of 799 non-DA WDs. The three black lines donote the best fitting of each model for non-DA WDs. }
\label{nonDA}
\end{figure}

\begin{figure}[!h]
\centerline {\includegraphics[width=0.9\textwidth]{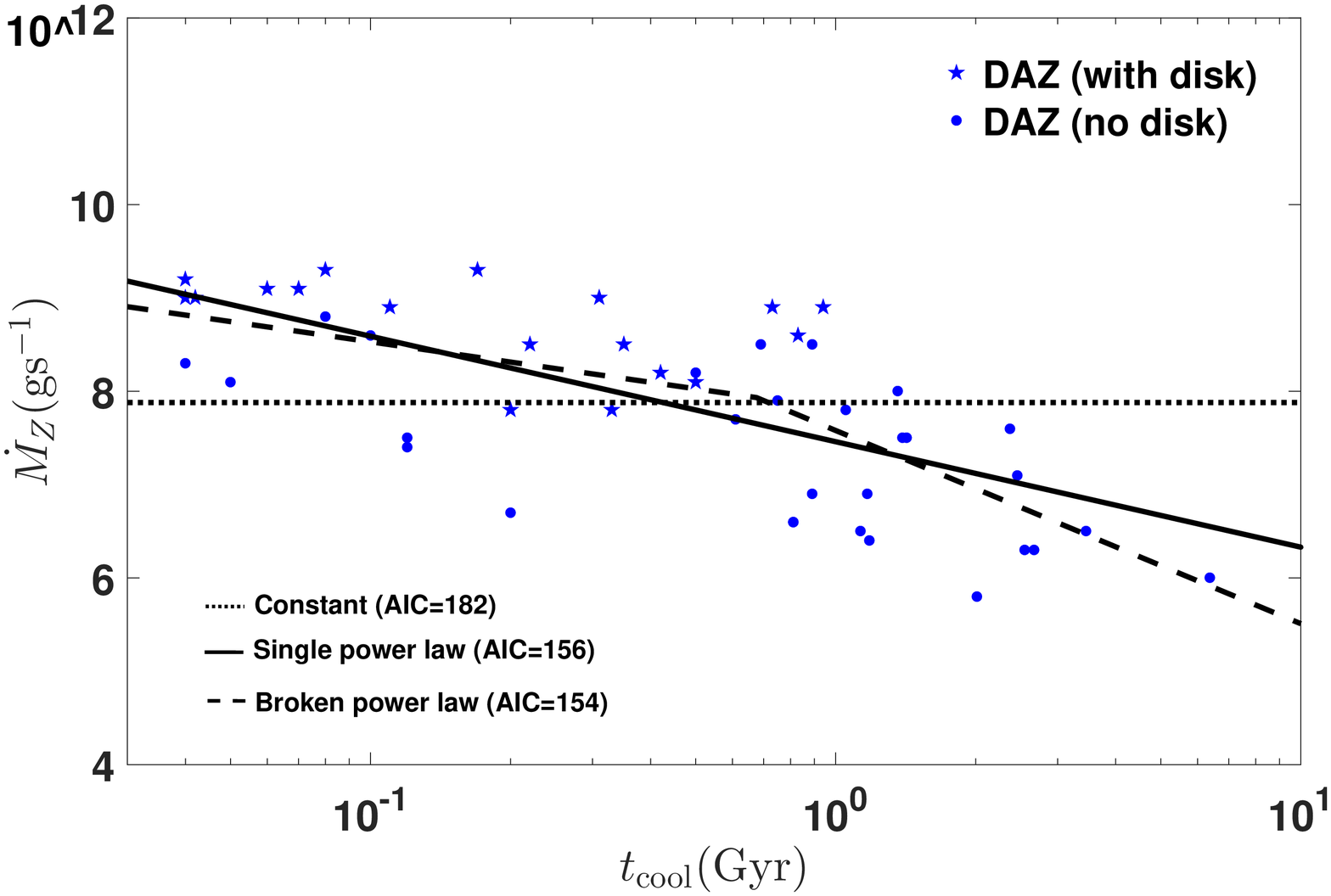}}
{Supplementary Figure 4: The mass accretion rate vs. the cooling ages of 47 single DAZ WDs. The three black lines donote the best fitting of each model for DA WDs. }
\label{DA}
\end{figure}

\begin{figure}[!h]
\centerline {\includegraphics[width=0.9\textwidth]{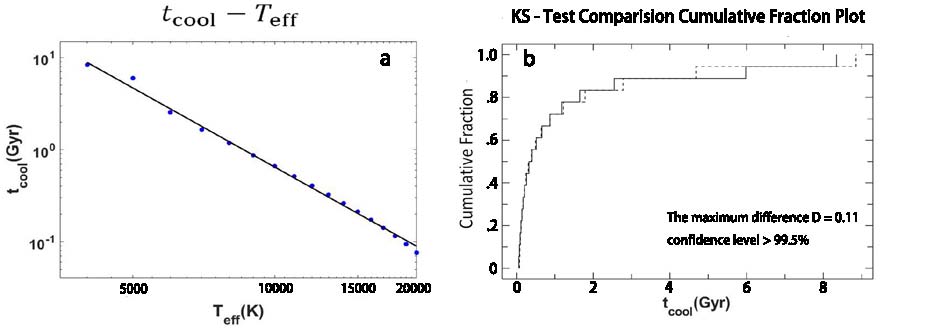}}
{Supplementary Figure 5: The comparison (left diagram) and the KS test result (right diagram) between $t_{\rm cool} (\rm Gyr)$ from Eqn. (\ref{eqS10}) and the data in the MWDD. In the diagram on the left, data in the MWDD is plot as solid blue points. The black line represents $t_{\rm cool} (\rm Gyr)$ from Eqn. (\ref{eqS10}).  }
\label{figS4}
\end{figure}

\begin{figure}[!h]
\centerline {\includegraphics[width=0.9\textwidth]{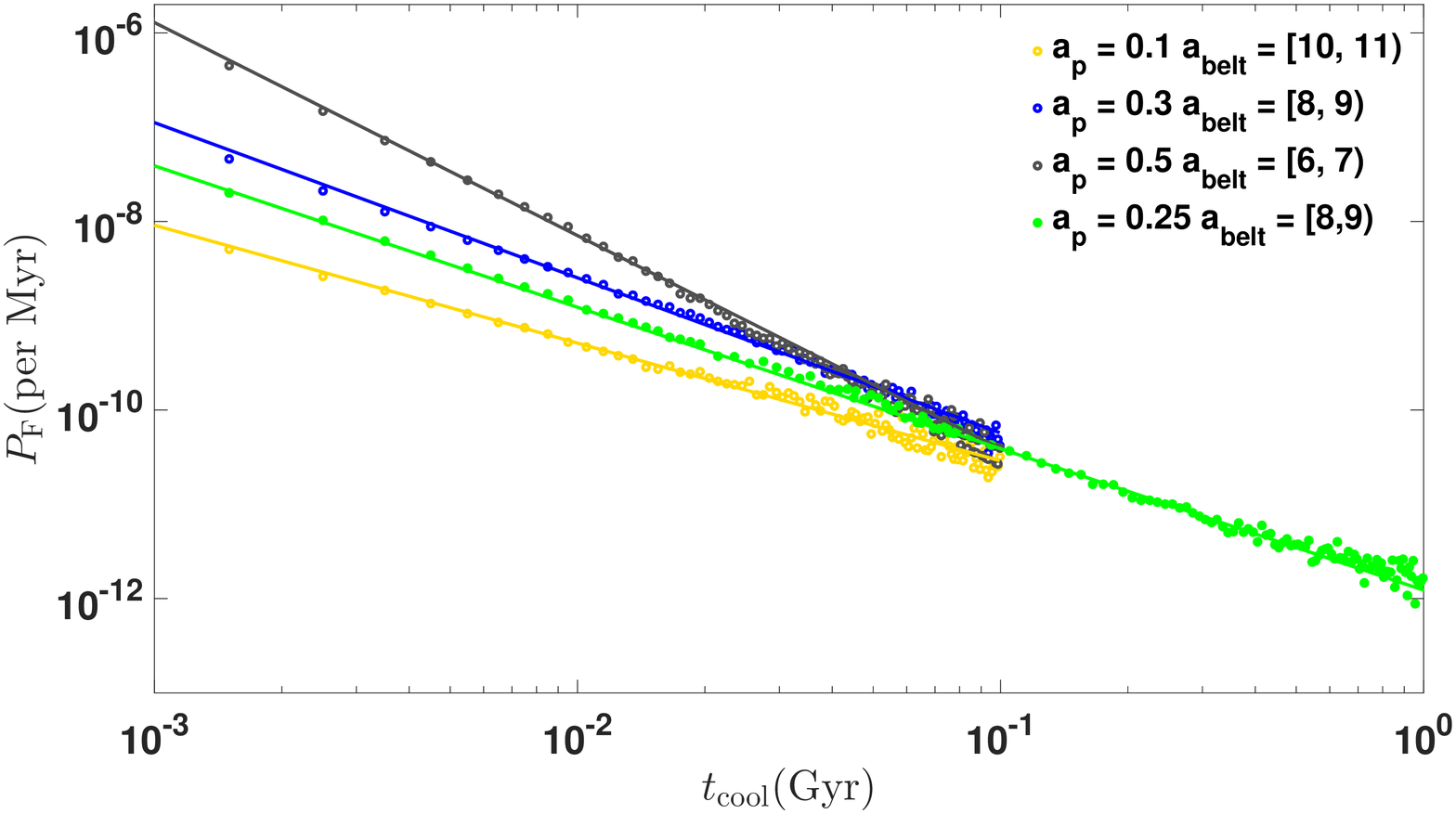}}
{Supplementary Figure 6: $P_{\rm F}$ as a function of the cooling ages for simulation results. The Jupter-mass planets are at 15.6AU. The different colours donote the simulation results for various parameters,  planet eccentricity ($e_{\rm p}$) and asteroid location ($a_{\rm belt}$). }
\label{figS5}
\end{figure}



%
%
%

\begin{figure}[!h]
\centerline {\includegraphics[width=0.9\textwidth]{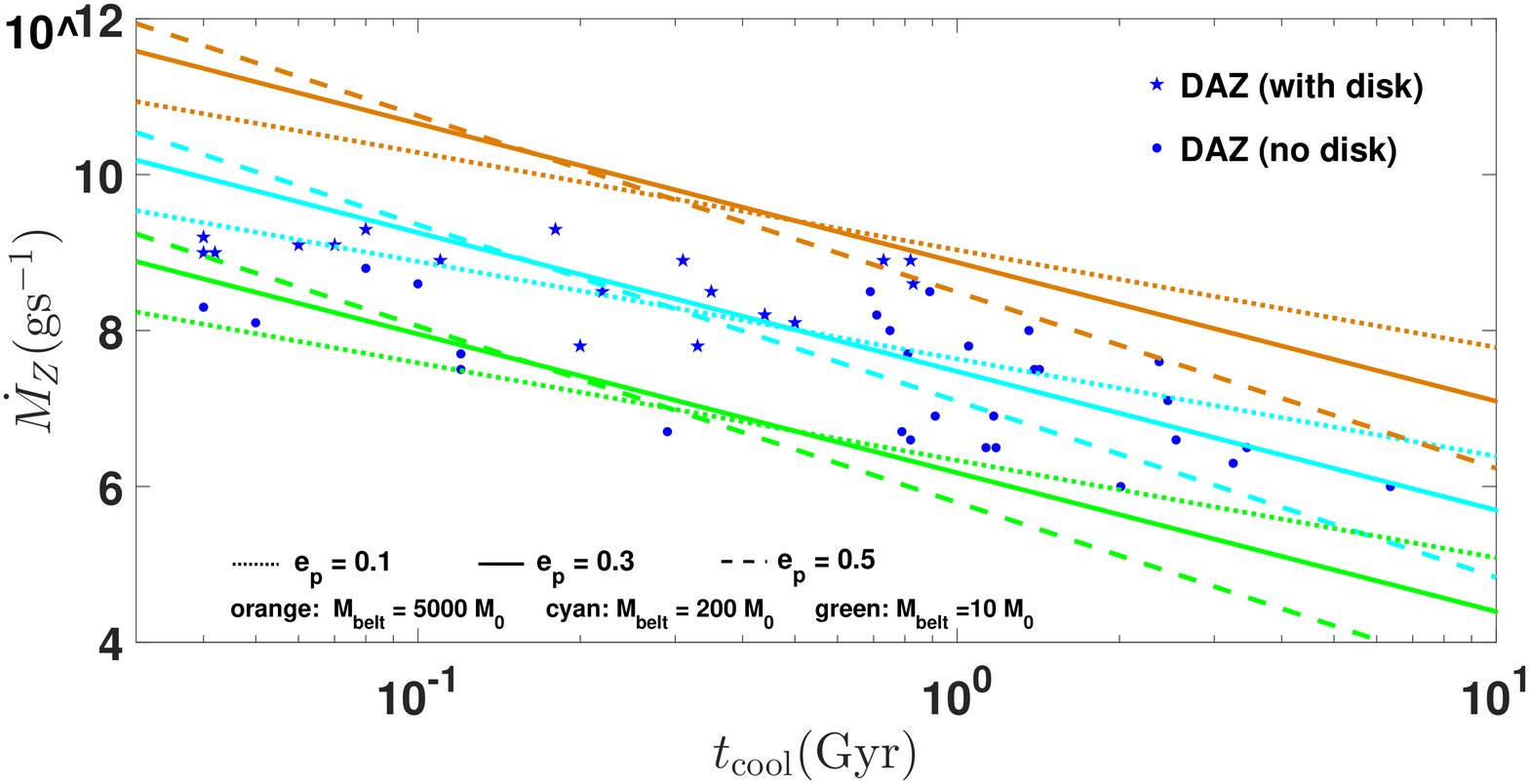}}
{Supplementary Figure 7: The compartion between the mass accretion rate and the dynamically falling rate $\dot{M}_{\rm DF}$ for different systems. DAZ WDs with disks are plotted as solid blue stars and those without disks are plotted as solid blue points. The  lines represent the caculated $\dot{M}_{\rm DF}$ for different initial conditions. }
\label{figS6}
\end{figure}

\begin{figure}[!h]
\centerline {\includegraphics[width=0.9\textwidth]{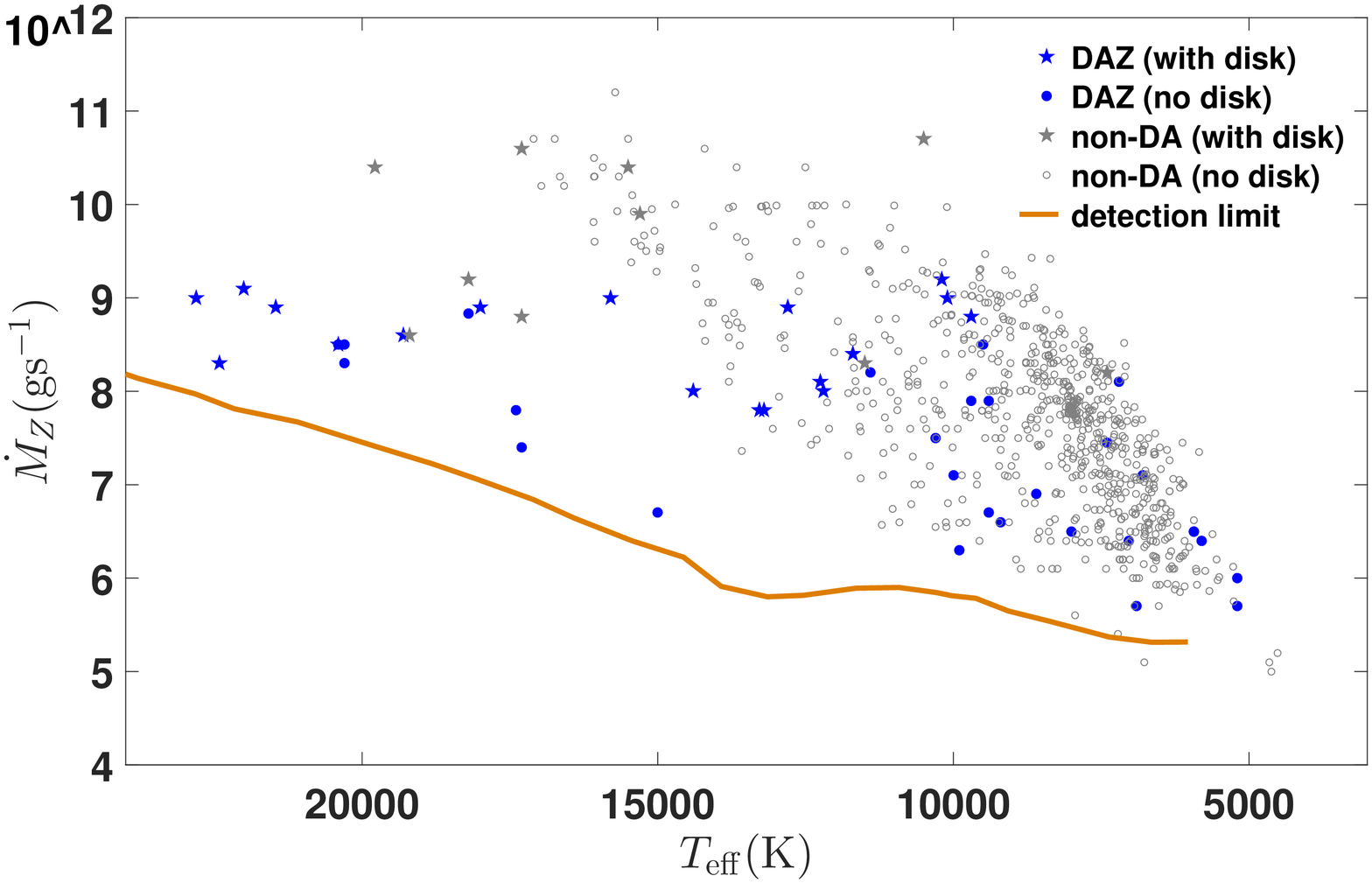}}
{Supplementary Figure 8: The mass accretion rate vs. the effective temperatures of all 846 WDs.  The solid orange line denotes the detction limit of the accretion rate for DA WDs.} 
\label{figS7}
\end{figure}

\begin{figure}[!h]
\centerline {\includegraphics[width=0.9\textwidth]{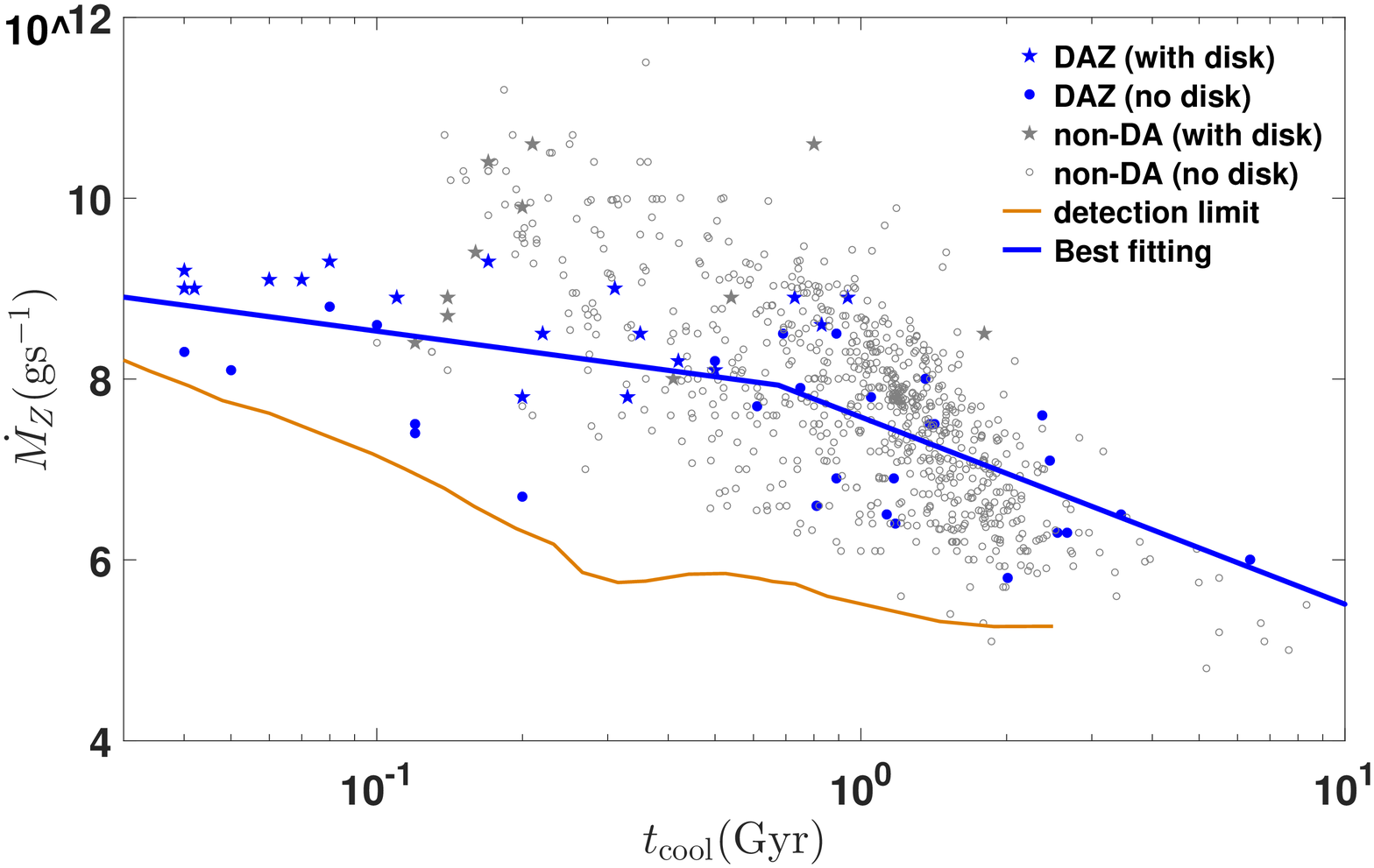}}
{Supplementary Figure 9: The mass accretion rate vs. the cooling ages of all 846 WDs. The solid orange line denotes the detction limit of the accretion rate for DA WDs. The best fitting line for DA WDs is plot as solid blue line.}
\label{figS8}
\end{figure}

\begin{figure}[!h]
\centerline {\includegraphics[width=0.9\textwidth]{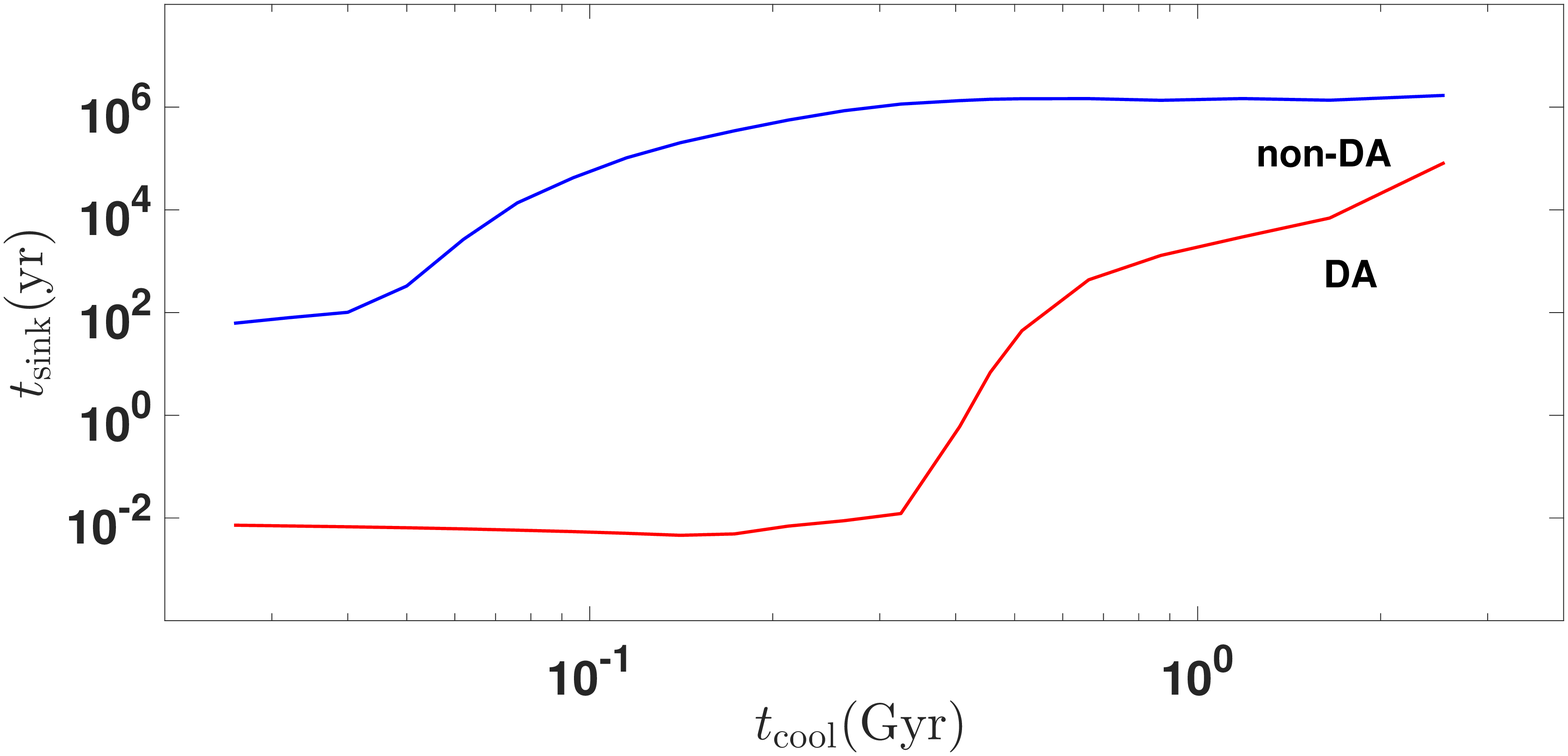}}
{Supplementary Figure 10: The metal sinking timescale vs. the cooling age of WDs both for Hydrogen-dominated atmospheres (DA) WDs (shown in red line) and Helium-dominated atmospheres (non-DA) WDs (shown in blue).}
\label{figS9}
\end{figure}

\clearpage

\begin{center}
\begin{table}[!h]
\centering
{\bfseries Supplementary Table 1:Metal-polluted WDs in our sample }
{\footnotesize
\label{TableS1}
\begin{tabular}{p{2.4cm}<{\centering}p{1.4cm}<{\centering}p{1.4cm}<{\centering}p{1.4cm}<{\centering}p{1.4cm}<{\centering}p{1.4cm}<{\centering}p{1.4cm}<{\centering}p{1.4cm}<{\centering}p{1.4cm}<{\centering}}
\hline
  Name &  SpT & $T_{\rm eff}({\rm K})$ & $\sigma T_{\rm eff}({\rm K})$ & $t_{\rm cool}({\rm Gyr})$ & $\sigma t_{\rm cool}({\rm Gyr})$ & $\log[\frac {\dot{M}_{\rm Z}}{\rm gs^{-1}}]$ & $\sigma\log[\frac {\dot{M}_{\rm Z}}{\rm gs^{-1}}]$ & Ref \\ \hline 
  WD $0002+729$   & DBZ & $13110$  & 338  & 0.31  & 0.03 & $7.7$  & 0.16  & $1$ \\
  WD $0032-175 $   & DAZ & $9200$  & 168  & 0.75 & 0.04 &  $6.6$  & 0.26  & $1$ \\ 
  WD $0033-114$   & DZ   & 7280    & 70   & 1.50 & 0.04 &  7.6    & 0.11  & $3$ \\
  WD $0046+051$   & DZ   & 6200    & 110  & 3.45 & 0.4  &  6.2    & 0.10  & $1$ \\
  WD $0047+190 $  & DAZ  & $16600$ & 674  & 0.11 & 0.02 &  $8.6$  & 0.22  & $2$ \\
\hline   
\multicolumn{9}{l}{
WDs are classified as D followed by A (hydrogen) and B (helium), with Z denoting metal lines. }\\
\multicolumn{9}{l}{
 d represents WD with detected circumstellar disk.}\\
\multicolumn{9}{l}{
DZ stars have helium-rich atmospheres but are too cool for He I absorption to be visible.}\\ 
\multicolumn{9}{l}{
$\dot{M}_{\rm Z}$ and $t_{\rm cool}$ represent the measured mass accretion rates and cooling ages of WDs.}\\
\multicolumn{9}{l}{
Reference: (1)  Farihi et al. 2009\cite{Farihi2009}; (2) Farihi et al. 2010\cite{Farihi2010b}; (3) Xu \& Jura 2012\cite{Xu2012} (4) Girven et al. (2012)\cite{Girven2012};} \\
\multicolumn{9}{l}{
(5) Bergfors et al. 2014)\cite{Bergfors2014}; (6) Dufour et al. 2007\cite{Dufour2007}; (7) Limoges et al. 2015\cite{Limoges2015}; }\\
\multicolumn{9}{l}{
(8) Koester \& Kelper 2015\cite{Koester2015}; (9) Kepler et al. 2015\cite{Kepler2015}; (10) Kepler et al. 2016\cite{Kepler2016}.}\\
\multicolumn{9}{l}{
A portion is shown here for guidance regarding its form and content. The entire table is available in a csv file. }\\

\end{tabular}}

\end{table}
\end{center}

\begin{center}
\begin{table}[!h]
\centering
\centering {\bfseries Supplementary Table 2: The fitting results of observation data}
{\footnotesize
\label{TableS2}
\begin{tabular}{c|p{2.5cm}<{\centering}p{2.2cm}<{\centering}p{1.6cm}<{\centering}p{2.2cm}<{\centering}p{1.6cm}<{\centering}p{2.2cm}<{\centering}p{1.6cm}<{\centering}}
\hline
   & \multicolumn{2}{c}{All (N=846)}   & \multicolumn{2}{c}{DA (N=47)}    & \multicolumn{2}{c}{non-DA (N=799)} \\ \hline
Model    & \multicolumn{2}{c}{}   & \multicolumn{2}{c}{AIC score}    & \multicolumn{2}{c}{} \\ \hline
Constant    &\multicolumn{2}{c}{5869}   & \multicolumn{2}{c}{182}    & \multicolumn{2}{c}{5506} \\
Single power law   & \multicolumn{2}{c}{5342}   & \multicolumn{2}{c}{156}    & \multicolumn{2}{c}{4920} \\
Broken power law   & \multicolumn{2}{c}{5265}   & \multicolumn{2}{c}{154}    & \multicolumn{2}{c}{4920} \\ \cline{2-7}
Coefficient      &     value    &    1 $\sigma$ interval &     value    &    1 $\sigma$ interval &     value    &    1 $\sigma$ interval \\ \cline{2-7}
 $\rm B_{0}$     &     7.76     &  (7.75, 7.77)  &   7.88  & (7.84, 7.92)     &   7.76     &  (7.75, 7.77) \\  
 $\rm A$         &     -2.25    &  (-2.25, -2.19) &  -1.13 & (-1.18, -1.08)  & -2.66      & (-2.69, -2.64) \\   
 $\rm B$         &     7.68     &   (7.67, 7.69)   &  7.46  & (7.43, 7.48)  &  7.71   & (7.70, 7.72) \\
 $\rm A_{1}$     &     1.43     &  (1.28, 1.89)   &  -0.72 & (-0.82, -0.53)  &  9.51 & (5.18, 12.39)  \\
 $\rm B_1  $     &     10.50    &  (10.36, 11.00) &  7.81  & (7.72, 8.01)    & 17.40 & (13.75, 19.92) \\
 $\rm A_2  $     &     -2.72    &  (-2.68, -2.77) &  -2.07 & (-2.29, -1.88)  & -2.81 & (-2.84,-2.78) \\
 $\rm B_2  $     &      7.72    &  (7.70, 7.72)   &  7.58  & (7.51,7.66)  & 7.73  & (7.71, 7.74) \\
 $\rm t_{12}(Gyr) $   &  0.21    &  (0.18, 0.22)   &  0.68 & (0.48, 0.93)  & 0.16  & (0.15,0.18)\\ \hline

\multicolumn{7}{l}{
1. Constant: Eqn. (6);   2. Single power law: Eqn. (7);  3. Broken power law: Eqn. (8).}\\
\end{tabular}}
\end{table}
\end{center}

\begin{center}
\begin{table}[!h]
\centering
{\bfseries Supplementary Table 3: Initial conditions and calculated coefficients of simulations }
{\footnotesize
\label{TableS3}
\begin{tabular}{cccccccc}
\hline
  & $e_{\rm p} $ &   &  $a_{\rm belt}$(AU) & $\alpha $  &  & $\beta $ & \\ \hline
  & $ 0.1 $  &    & [9, 10) & $ -1.25(0.08) $ &  &  $-11.79(0.16)$  & \\
  & $ 0.1  $ &    & [10, 11)& $ -1.52(0.11) $ &  &  $-12.70(0.23)$  &  \\
  & $ 0.1 $ &     & [11, 12) & $-2.34(0.06) $ &  &  $-15.06(0.18)$  &  \\
  & $ 0.2 $ &     & [8, 9) & $ -1.42(0.15)  $ &  &  $-12.17(0.16)$  &  \\
  & $ 0.2 $ &     & [9, 10) & $-1.68(0.06)  $ &  &  $-12.89(0.10)$  &  \\
  & $ 0.2 $ &     & [10, 11) & $-1.92(0.06)$ &  &   $-13.47(0.16)$ &  \\
  & $ 0.25 $ &    & [8, 9) & $-1.50(0.20)$ &  &  $-11.91(0.14)$ &  \\
  & $ 0.3 $ &    & [8, 9) & $ -1.78(0.07) $ &  &  $-12.03(0.08)$ &  \\
  & $ 0.3 $ &    & [9, 10) & $ -2.11(0.11) $ &  &  $-13.47(0.20)$ &  \\
  & $ 0.4 $ &    & [7, 8) & $ -1.57(0.06)  $ &  &  $-11.37(0.13)$ &  \\
  & $ 0.4 $ &    & [8, 9) & $ -2.68(0.18) $ &  &  $-14.52(0.22)$ & \\
  & $ 0.5 $ &    & [6, 7) & $ -2.26(0.13) $ &  &  $-12.63(0.25)$ &  \\ \hline 
\multicolumn{8}{l}{
$e_{\rm p}$ is the orbital eccentricity of the Jovian planet.}\\
\multicolumn{8}{l}{
$a_{\rm belt}$ is the range of the asteroid belt.}\\
\multicolumn{8}{l}{
$\alpha$ and $\beta$ are the coefficients of the power law decay.}\\
\end{tabular}}

\end{table}
\end{center}

\end{document}